\documentclass[letterpaper,journal]{IEEEtran}

\usepackage{iftex}
\ifPDFTeX
  \usepackage[T1]{fontenc}
  \usepackage[utf8]{inputenc}
\else
  \usepackage{fontspec}
\fi
\usepackage{amsmath,amsfonts}
\usepackage{algorithmic}
\usepackage{algorithm}
\usepackage{array}
\usepackage[caption=false,font=normalsize,labelfont=sf,textfont=sf]{subfig}
\usepackage{textcomp}
\usepackage{stfloats}
\usepackage{url}
\usepackage{verbatim}
\usepackage{graphicx}
\usepackage{cite}
\usepackage{booktabs}
\usepackage{calc}
\usepackage{capt-of}
\usepackage{placeins}
\usepackage{hyperref}
\hypersetup{hidelinks}

\providecommand{\real}[1]{#1}

\begin{document}

\title{Structured Neural Modeling of Daily Arctic Sea-Ice Concentration Evolution: Physical-Trajectory-Driven Learning and Forecast-Domain Adaptation}

\author{Maqun Zhang$^1$, Feng Gao$^1$, Wankun Chen$^1$, Hui Yu$^2$, Yanhai Gan$^{1,*}$, and Junyu Dong$^{1,*}$%
\thanks{This work was supported by the Key R\&D Program of Shandong Province under Grant 2025CXPT185; the Leverhulme Trust through Project VP1-2020-044; the National Natural Science Foundation of China under Grant 42406192; the Fundamental Research Funds for the Central Universities under Grants 202413040 and 202572015; the National Science and Technology Major Project of China under Grant 2022ZD0117201; and the Postdoctoral Project of Qingdao under Grant QDBSH20240102021.}%
\thanks{Corresponding authors: Yanhai Gan (ganyanhai@ouc.edu.cn) and Junyu Dong (dongjunyu@ouc.edu.cn).}%
\thanks{Maqun Zhang, Feng Gao, Wankun Chen, Yanhai Gan, and Junyu Dong are with the State Key Laboratory of Physical Oceanography and Faculty of Information Science and Engineering, Ocean University of China, Qingdao 266100, China. Hui Yu is with the School of Psychology and Neuroscience, University of Glasgow, Glasgow G12 8QQ, U.K.}}

\markboth{IEEE Transactions on Geoscience and Remote Sensing}%
{Structured Neural Modeling of Daily Arctic Sea-Ice Concentration Evolution}

\maketitle

\begin{abstract}
Accurate modeling of the daily evolution of sea ice concentration (SIC) is central to improving the credibility and operational forecasting capability of deep learning-based sea ice prediction. However, existing deep learning methods often couple the underlying sea ice evolution relationships and data errors within high-dimensional nonlinear mappings, making it difficult to construct a stable and verifiable evolution core and apply it reliably to practical forecasting. To address this issue, this study proposes a reanalysis-forecast dual-domain decoupled framework for learning sea ice evolution operators. The framework builds upon a lightweight physical baseline to generate daily evolution trajectories, employs a temporally constrained joint multi-lead compensation network to compensate for unresolved processes, and introduces an ice-mass-aware transport mechanism to suppress numerical dissipation. In the forecasting stage, the parameters of the base evolution core are fixed, while a lightweight variable-semantic adaptation mechanism calibrates inter-domain distributions and evolution responses, thereby separating forecast-domain errors from base evolution errors. Experiments show that the constructed base evolution core can accurately and stably simulate daily sea ice evolution at both short-term and annual scales under reanalysis forcing, and can be effectively transferred to the forecast domain through lightweight adaptation, achieving stable practical forecasting capability while preserving the base evolution structure. The source code will be made publicly available at
\url{https://github.com/zhangmaqun65535/SNM}
upon acceptance of this manuscript.
\end{abstract}

\begin{IEEEkeywords}
sea ice concentration, evolution operator, physics-guided deep learning, temporally constrained joint learning, forecast forcing, domain adaptation.
\end{IEEEkeywords}

\section{Introduction}

Changes in Arctic sea ice have important implications for the climate system and polar activities, and accurate prediction of daily sea ice concentration (SIC) variations is of substantial scientific and practical value \cite{ref1,ref2,ref3}. Deep learning can extract complex nonlinear spatiotemporal relationships from multi-source historical data and has become an important technical route for sea ice prediction \cite{ref4,ref5}. However, deep learning models typically embed sea ice evolution processes within high-dimensional parameters and nonlinear mappings, making it difficult to determine whether they have learned stable and credible daily evolution relationships or to verify how their predictions are formed \cite{ref6,ref7}. Therefore, constructing a daily SIC evolution model with an explicit evolution structure and verifiable process roles, while retaining the data representation capability of deep learning, is essential for improving the scientific credibility and practical forecasting capability of deep learning-based sea ice prediction.

Existing sea ice simulation and prediction methods mainly fall into three categories: numerical methods, data-driven methods, and physics-guided methods \cite{ref4,ref8,ref9,ref10,ref11}. Numerical methods have explicit process representations \cite{ref8}, but their fixed parameterizations make it difficult to learn unresolved processes from large-scale data \cite{ref10,ref11}, and finite-resolution transport may introduce numerical dissipation \cite{ref9}. Data-driven methods have strong nonlinear representation capabilities, but the learned evolution relationships are embedded within the network and are difficult to verify \cite{ref4,ref5,ref6,ref7}. Physics-guided methods combine the strengths of these two approaches \cite{ref10,ref11,ref12}, but the functional boundary between physical evolution and neural correction remains unclear, and errors in the physical baseline may also propagate to the output \cite{ref12}; moreover, the structural difference between daily recurrence and joint multi-lead prediction limits computational scalability \cite{ref13,ref14}. Therefore, existing methods have not yet simultaneously achieved an explicit evolution structure, sufficient data learning capacity, and verifiable process decomposition.

These limitations indicate that the key to physics-guided sea ice modeling is not a simple superposition of numerical prediction and neural networks, but a clear definition of their functional boundaries in daily evolution. Physical propagation should provide a stable evolution backbone and control numerical dissipation, while neural networks should compensate for unresolved processes within clearly defined boundaries; meanwhile, multi-lead learning should achieve efficient joint modeling under temporal constraints. Therefore, it is necessary to construct a unified sea ice evolution model structured around physical evolution as the backbone and constrained data-driven compensation as the correction, thereby achieving both explicit daily evolution representation and efficient joint multi-lead learning capability.

Reliable sea ice prediction requires accurate modeling of the daily evolution operator and, in practical forecasting, separation of evolution errors from forecast-domain errors. To this end, this study learns a base evolution core from reanalysis data, in which a daily recurrent lightweight physical baseline represents the main evolution processes, and a temporally constrained joint multi-lead compensation network compensates for unresolved processes and baseline errors. In the forecasting stage, to prevent observational and forecast input errors from being absorbed into the learned evolution structure, this core is kept frozen, and only independent fine-tuning modules are introduced to adapt input-domain differences and correct error responses, thereby enabling the stable sea ice evolution core to support practical forecasting.

The main contributions of this study are summarized as follows:

\begin{itemize}
\item A reanalysis-forecast dual-domain decoupled framework is proposed for deep learning-based sea ice evolution operators. The framework constructs a physically structured evolution core from reanalysis data and models forecast-domain errors through independent adaptation layers, thereby separating evolution relationships from input errors.

\item A low-cost physical-trajectory-driven temporally constrained joint multi-lead learning framework is designed for constructing the evolution core. To address the structural difference between daily recurrence and joint multi-lead learning, low-cost physical trajectories explicitly carry state evolution, while multi-lead compensation terms are jointly estimated under temporal constraints, thereby coordinating physical recurrence with efficient joint learning.

\item A variable-semantic adaptation mechanism is proposed for reanalysis-forecast domain differences. To address input distribution shifts and forecast errors, adaptation is decomposed into inter-domain distribution calibration and evolution-response correction according to the physical semantics and error structures of the variables; lightweight, independently parameterized adaptation mappings then separate forecast-domain errors from base evolution errors.

\item An ice-mass-aware transport mechanism is proposed for sea ice dynamics. To address numerical dissipation and total ice-mass loss during multi-step dynamic propagation, spatial redistribution is constrained by total ice-mass conservation, while unresolved ice mass is retained through recursive relative-ice-state representation, improving ice-mass conservation and suppressing systematic negative drift.
\end{itemize}

\section{Related Work}

\subsection{Data-Driven Simulation and Prediction of SIC}

With the accumulation of long-term satellite observations and reanalysis products, data-driven methods have increasingly been applied to gridded SIC field modeling, and early studies demonstrated the ability of deep networks to learn nonlinear spatiotemporal relationships from sea ice states and environmental variables \cite{ref15,ref16}. Subsequently, these models progressed from monthly-to-seasonal scales to daily and multi-lead prediction, with recurrent states, multi-time inputs, and multi-scale features enhancing temporal-dependence modeling \cite{ref4,ref17,ref18,ref19,ref20,ref21,ref22}. Recent studies have placed greater emphasis on available information and spatial details: multi-task networks jointly model SIC and ice velocity \cite{ref23}, frequency-domain compensation and generative priors improve ice-edge and high-resolution representations \cite{ref5,ref24}, and the integration of operational atmospheric forecasts, ice charts, and satellite observations has advanced kilometer-scale short-range sea ice prediction \cite{ref3}. Existing methods have improved multi-source fusion, multi-lead prediction, and spatial-detail representation for SIC, and some studies can also quantify variable contributions. However, state evolution, forcing responses, and data errors remain insufficiently separated in a structured manner, making stable daily evolution relationships difficult to verify. Therefore, explicit physical processes and conservation constraints remain necessary.

\subsection{Physics-Guided Spatiotemporal Evolution and Conservation Modeling}

To improve the physical consistency and evolution stability of data-driven models, physics-guided methods incorporate governing equations or conservation relationships into training objectives and network architectures, or advance the state with physical models and correct errors through neural networks \cite{ref6,ref11,ref25,ref26}. Recently, the integration of numerical dynamical cores and learned corrections has been adopted in Earth system modeling and has further developed into hybrid frameworks in which low-resolution dynamical cores generate physical trajectories and neural operators correct systematic biases \cite{ref10,ref27,ref28}. In the sea ice domain, recent studies have begun to jointly model SIC and ice velocity and to correct state-dependent errors in physical sea ice models through learned modules \cite{ref12,ref29}; conservation-based sea ice simulators further improve long-term stability and process attribution through explicit mass and area budgets \cite{ref30}. However, existing studies have not yet unified low-cost physical propagation, independent neural compensation, and transport-induced ice-mass-loss control in daily SIC modeling, and the functional boundary between the physical evolution backbone and data-driven correction remains to be further clarified.

\subsection{Multi-Lead Prediction and Temporal-Dependence Modeling}

Multi-lead prediction requires temporal dependence to be preserved while controlling recursive errors and computational cost. Recurrent models such as ConvLSTM and PredRNN maintain temporal order through recurrent states, but their serial dependence limits parallel efficiency, and single-step errors accumulate as the forecast advances; related training strategies can only partially alleviate this issue \cite{ref31,ref32,ref33}. Joint prediction models such as SimVP, Earthformer, and TFT can directly generate future sequences \cite{ref13,ref34,ref35}, and recent studies have further achieved synchronous multi-lead output and non-recurrent full-trajectory prediction \cite{ref14,ref36}. Causal convolutions and masked attention can restrict the information visibility of different output time positions, making temporal constraints compatible with joint multi-lead computation \cite{ref37,ref38,ref39}. However, existing methods mainly perform direct recurrence or joint prediction of complete states, while limited attention has been given to carrying state evolution through recurrent physical trajectories and jointly estimating multi-lead compensation terms under temporal constraints.

\subsection{Forecast-Domain Transfer and Error Adaptation}

When models trained on reanalysis data are transferred to practical forecasting, sea ice initial conditions and external forcing are provided by observed SIC and forecast fields, respectively, leading to input distribution shifts and observation and forecast errors from different sources. Classical domain adaptation methods reduce overall discrepancies by learning domain-invariant features or aligning feature statistics \cite{ref40,ref41}, while Residual Adapter and FiLM adjust shared backbones through a small number of additional parameters \cite{ref42,ref43}. Recently, WeatherPEFT further extends parameter-efficient fine-tuning to task adaptation in weather models \cite{ref44}. In forecasting systems, statistical post-processing and the integration of data assimilation with machine learning have been applied to correct systematic biases and state-dependent errors \cite{ref45,ref46,ref47,ref48}; TIFNet is trained on reanalysis data and fine-tuned with operational forecast data \cite{ref14}, ReST corrects regional and season-dependent biases through spatiotemporal conditional modulation \cite{ref49}, and related ideas have also been applied to sea ice model transfer and online SIC bias correction \cite{ref12,ref50}. However, existing methods mainly target overall distribution alignment or final prediction-bias correction, with limited attention to lightweight adaptation of errors in the observation and forecast domains according to the physical semantics of different variables under a fixed sea ice evolution core.

\section{Methods}

This study first learns a daily sea ice evolution core in the reanalysis domain, then freezes the core and adapts the observational initial state and forecast forcing to separate sea ice evolution relationships from forecast-input errors.

\begin{figure*}[!t]
\centering
\includegraphics[width=\textwidth]{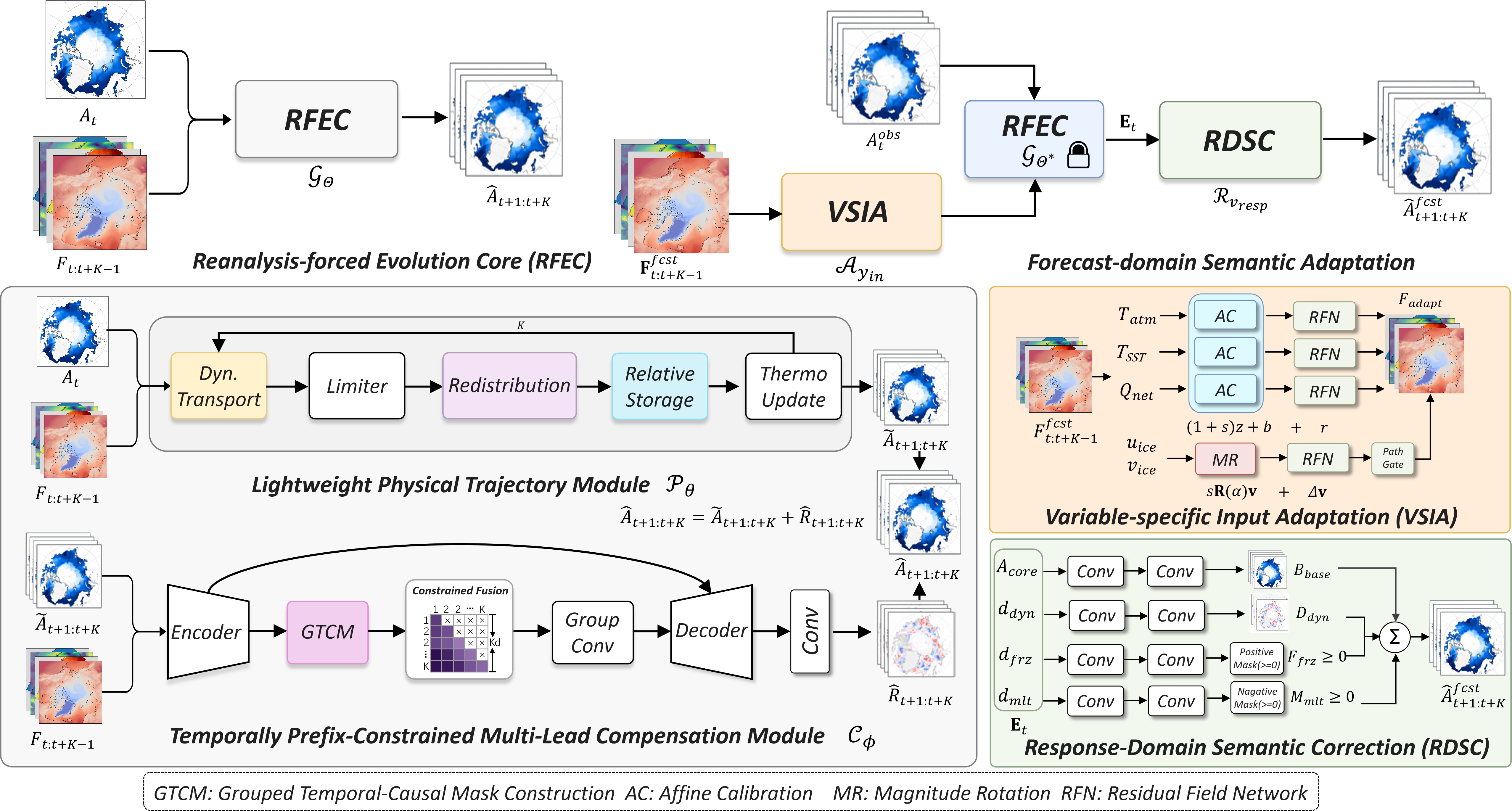}
\caption{Overall architecture of the proposed network. In the reanalysis domain, the model takes the initial SIC and multi-lead forcings as inputs, generates the daily state trajectory through the Lightweight Physical Trajectory Module ($P_\theta$), and jointly estimates residuals in a single forward pass using the Temporally Prefix-Constrained Multi-Lead Compensation Module ($C_\phi$), thereby producing the evolution-core output. During forecasting, the evolution core remains frozen. Variable-Specific Input Adaptation (VSIA) calibrates the observed initial condition and forecast forcings, while Response-Domain Semantic Correction (RDSC) further refines the output. The model thereby establishes a structured separation among physical propagation, multi-lead residual learning, and forecast-domain adaptation.}
\label{fig:overall-framework}
\end{figure*}

The overall architecture is shown in Fig.~\ref{fig:overall-framework}.

\subsection{Overall Task and Problem Formulation}

This section defines the overall task at an abstract level, while specific module implementations are detailed in the following sections.

Let \(A_t\) denote the SIC field on day \(t\), and let \(\mathbf F_t\) denote the corresponding external forcing fields. The daily evolution of sea ice can be represented by an unknown evolution operator:

\[
A_{t+1}=\mathcal G(A_t,\mathbf F_t).
\tag{1}
\]

The primary objective of this study is to approximate the daily sea ice evolution operator \(\mathcal G\). Accordingly, the physical baseline \(\mathcal P_{\theta}\) represents the principal dynamical transport and thermodynamic changes, while the compensation operator \(\mathcal C_{\phi}\) learns unresolved processes and model errors:

\[
\widetilde A_{t+1}=\mathcal P_{\theta}(A_t,\mathbf F_t).
\tag{2}
\]

The complete model corrects the physical-baseline estimate through the compensation operator:

\[
\widehat A_{t+1}
=
\Pi_{[0,1]}
\left[
\widetilde A_{t+1}
+
\mathcal C_{\phi}
(A_t,\mathbf F_t,\widetilde A_{t+1})
\right].
\tag{3}
\]

Here, \(\widetilde A_{t+1}\) and \(\widehat A_{t+1}\) denote the physical-baseline estimate and complete-model output, respectively, while \(\Pi_{[0,1]}\) enforces the physically admissible range of SIC.

The base evolution model learns sea ice evolution relationships from high-quality reanalysis states and forcing fields, whereas practical forecasting provides only the observational initial state \(A_t^{\mathrm{obs}}\) and error-prone forecast forcing \(\mathbf F_t^{\mathrm{fcst}}\). Therefore, the complete evolution model \(\widehat{\mathcal G}_{\theta^*,\phi^*}\), learned in the reanalysis domain, is kept fixed, while a state adaptation operator \(\mathcal A_{\psi_A}^{A}\) and a forcing adaptation operator \(\mathcal A_{\psi_F}^{F}\) are introduced. The one-step forecasting task is formulated as

\[
\widehat A_{t+1}^{\mathrm{fcst}}
=
\widehat{\mathcal G}_{\theta^*,\phi^*}
\left(
\mathcal A_{\psi_A}^{A}\!\left(A_t^{\mathrm{obs}}\right),
\mathcal A_{\psi_F}^{F}\!\left(\mathbf F_t^{\mathrm{fcst}}\right)
\right).
\tag{4}
\]

Subscripts distinguish learnable parameters from different sources: \(\theta\) and \(\phi\) denote the parameters of the physical baseline and compensation network in the reanalysis domain, respectively, while \(\psi\) collectively denotes parameters introduced in the forecast domain; the superscript \(^*\) indicates parameters that have been trained and remain fixed during forecast-domain adaptation. In the state adaptation operator \(\mathcal A_{\psi_A}^{A}\) and forcing adaptation operator \(\mathcal A_{\psi_F}^{F}\), the subscripts \(A\) and \(F\) identify the state and forcing semantics, respectively; \(\psi_F\) collectively denotes the forcing-adaptation parameters, whose decomposition is detailed in Section III-C. Forecast-domain adaptation optimizes only the newly introduced parameters denoted by \(\psi\), thereby preserving the sea ice evolution relationships learned in the reanalysis domain and separating observational errors, forecast-forcing errors, and base-simulation errors.

\subsection{Physics-Guided Temporally Constrained Joint Multi-Lead Evolution Framework}

\subsubsection{Physical-Trajectory-Driven Joint Multi-Lead Compensation Framework}

To recursively extend the daily evolution relationship defined in Section III-A to multi-lead simulation, the model must preserve the evolution dependency between adjacent time steps while exploiting sequence-prefix information for joint multi-lead computation. Accordingly, this study constructs an evolution framework comprising a lightweight physical baseline and a joint multi-lead compensation network.

This section considers the reanalysis-domain evolution model, with \(\mathbf F\) denoting reanalysis forcing fields throughout. For a \(K\)-step simulation, the lightweight physical baseline advances daily according to Eq. (2), yielding the physical trajectory \(\widetilde{\mathbf A}_{t+1:t+K}\). Here, \(\mathbf F\) comprises vector sea ice velocity, sea surface temperature, 2-m air temperature, and net heat flux, from which the physical baseline represents the principal dynamical transport and thermodynamic changes. Dynamical transport and conservation correction are governed by deterministic operators, whereas the learnable parameters \(\theta\) are confined to the thermodynamic-change network.

Rather than learning complete sea ice evolution, the compensation network directly processes an explicit multi-lead sequence comprising the physical trajectory and external forcing. Through a single forward pass, the network jointly estimates the conditional residuals for \(K\) forecast leads:

\[
\begin{aligned}
\widehat{\mathbf R}_{t+1:t+K}
&=
\mathcal C_{\phi}
\left(
\widetilde{\mathbf A}_{t+1:t+K},
\mathbf F_{t:t+K-1}
\right),\\
\widehat{\mathbf A}_{t+1:t+K}
&=
\Pi_{[0,1]}
\left[
\widetilde{\mathbf A}_{t+1:t+K}
+
\widehat{\mathbf R}_{t+1:t+K}
\right].
\end{aligned}
\tag{5}
\]

The input at the \(k\)-th time step is denoted by \(X_{t+k}=[\widetilde A_{t+k},\mathbf F_{t+k-1}]\), which pairs the target-day physical state with the preceding-day forcing that drives the corresponding state transition. The initial state \(A_t\) has already been propagated stepwise into the physical trajectory, while the current and preceding sequence features jointly provide the recent evolution context.

This structure preserves the daily evolution relationship through the sequentially generated physical trajectory while reformulating the principal learnable computation as joint multi-lead residual estimation. The compensation network directly processes explicitly organized physical-state and forcing sequences rather than compressing them entirely into a recurrent hidden state; the temporal constraints on the joint output are described in the next subsection.

For the network formulation, the multi-step input in Eq. (5) is denoted by \(\mathbf X_{t+1:t+K}=(X_{t+1},\ldots,X_{t+K})\). The joint multi-lead compensation network comprises a shared spatial encoder \(\mathcal E_{\phi_E}\), a sequence feature fusion module \(\mathcal T_{\phi_T}\), and a shared spatial decoder \(\mathcal D_{\phi_D}\), with the processing pipeline summarized as

\[
\mathbf X_{t+1:t+K}
\xrightarrow{\mathcal E_{\phi_E}}
\mathbf Z_{t+1:t+K}
\xrightarrow{\mathcal T_{\phi_T}}
\mathbf H_{t+1:t+K}
\xrightarrow{\mathcal D_{\phi_D}}
\widehat{\mathbf R}_{t+1:t+K}.
\tag{6}
\]

Here, \(X_{t+k}\), \(Z_{t+k}\), and \(H_{t+k}\) denote the input, encoded feature, and fused feature at an individual time step, respectively, while the corresponding bold symbols denote sequences of \(K\) time steps. The learnable parameters of the compensation network are denoted by \(\phi=(\phi_E,\phi_T,\phi_D)\). The spatial encoder extracts the spatial representation at each time step, the sequence feature fusion module integrates multi-step evolution information, and the spatial decoder maps the fused features to the corresponding SIC compensation residuals. The organization and prefix-constrained temporal interactions of the multi-step features are described in Section III-B2; the layer-wise architecture of the compensation network is provided in Supplementary Materials C.1--C.3, the principal structure of the physical baseline in Supplementary Materials D.1--D.4, and the training configuration in Supplementary Materials A.3.

\subsubsection{Temporal Constraints on Joint Multi-Step Outputs}

The joint multi-lead compensation network simultaneously receives physical states and forcing information from multiple time steps. If cross-temporal features are mixed without restriction, the \(k\)-th output may incorporate physical states and forcing fields from later time steps, resulting in future information leakage. Accordingly, a lower-triangular temporal-prefix constraint is imposed on cross-temporal information exchange.

As described in Section III-B1, the spatial encoder and decoder operate separately at each time step, while cross-temporal information is exchanged only within the sequence feature fusion module \(\mathcal T_{\phi_T}\). Let \(Z_{t+k}\) and \(H_{t+k}\) denote the spatial representations on day \(t+k\) before and after feature fusion, respectively, where \(t\) is the simulation initialization day and \(j,k\in\{1,\ldots,K\}\) are relative time-step indices. Sequence feature fusion is formulated as

\[
\begin{bmatrix}
H_{t+1}\\
H_{t+2}\\
\vdots\\
H_{t+K}
\end{bmatrix}
=
\begin{bmatrix}
W_{1,1} & 0 & \cdots & 0\\
W_{2,1} & W_{2,2} & \cdots & 0\\
\vdots & \vdots & \ddots & \vdots\\
W_{K,1} & W_{K,2} & \cdots & W_{K,K}
\end{bmatrix}
\begin{bmatrix}
Z_{t+1}\\
Z_{t+2}\\
\vdots\\
Z_{t+K}
\end{bmatrix}.
\tag{7}
\]

Here, \(W_{k,j}\) is the learnable mapping from the \(j\)-th temporal channel group to the \(k\)-th temporal channel group and belongs to the sequence-fusion parameters \(\phi_T\); the upper-triangular entries of the matrix are fixed at zero. Consequently, the \(k\)-th time step can receive features only from time steps \(1\) through \(k\) and cannot access subsequent physical states or forcing information. The detailed weight structure is provided in Supplementary Materials C.

This constraint fixes the visibility relationships among temporal channels rather than the specific weights of the retained connections. Feature fusion for all time steps is still completed within a single masked tensor operation, thereby preserving joint output through one forward pass.

When only the first \(L<K\) steps are required, no connections exist between the first \(L\) rows of the matrix and inputs \(L+1\) through \(K\). Consequently, subsequent temporal channels may be padded with arbitrary placeholder values without affecting the first \(L\) outputs.

Thus, the temporal constraint prevents future information leakage and ensures that unavailable subsequent forcing does not contaminate outputs for the available forecast steps.

\subsubsection{Area-Constrained Redistribution and Relative Ice-Mass Representation}

The physical trajectory described in Section III-B1 requires continuous rollout, and its stability directly affects subsequent compensation. The baseline dynamical structure is described in Supplementary Materials D. Interpolative transport may reduce the regional integral of SIC; meanwhile, velocity-field convergence may raise local SIC above its physical upper bound, and direct clipping discards compression that cannot be represented explicitly. These errors accumulate during rollout, producing numerical dissipation and systematic negative drift.

Accordingly, area-constrained redistribution and relative ice-mass storage are introduced. The former reallocates the equivalent sea ice area deficit after transport to ice-covered regions with local loss and remaining capacity; the latter stores overflow generated by upper-bound clipping in a relative ice-mass state and releases it when capacity subsequently becomes available. The relative ice-mass state retains unresolved compression information but does not represent actual sea ice thickness.

The dynamical state is augmented from \(A_t\) to \((A_t,S_t)\), with its update summarized as

\[
\begin{aligned}
\left(A_{t+1}^{D},S_{t+1}\right)
&=
\left(
\mathcal R_{\mathrm{rel}}
\circ\mathcal R_{\mathrm{area}}
\circ\mathcal P_{\mathrm{dyn}}^{0}
\right)\\
&\quad\left(A_t,S_t,\mathbf v_t\right),\\
0\leq A_{t+1}^{D}&\leq1,\qquad S_{t+1}\geq0.
\end{aligned}
\tag{8}
\]

Here, \(\mathbf v_t\) denotes vector sea ice velocity, \(\mathcal P_{\mathrm{dyn}}^{0}\) denotes the deterministic dynamical transport within the physical baseline \(\mathcal P_{\theta}\), and \(\mathcal R_{\mathrm{area}}\) and \(\mathcal R_{\mathrm{rel}}\) denote area-deficit correction and relative ice-mass updating, respectively. All three are deterministic operators and introduce no additional trainable parameters. The two correction mechanisms address area loss at the current time step and compression-information loss across time steps, respectively, thereby improving physical-trajectory stability during continuous rollout. The update procedure is detailed in Algorithm~\ref{alg:area-relative-update}.

\begin{table*}[!t]
\centering
\scriptsize
\refstepcounter{algorithm}
\textbf{Algorithm~\thealgorithm.} Area-Constrained Redistribution and Relative Ice-Mass Update\par
\label{alg:area-relative-update}
\vspace{2pt}
\begin{tabular}{@{}>{\raggedright\arraybackslash}p{0.19\textwidth}>{\raggedright\arraybackslash}p{0.75\textwidth}@{}}
\toprule
\textbf{Item} & \textbf{Description} \\
\midrule
\textbf{Input} & Current SIC \(A_t\), relative ice mass \(S_t\), vector sea ice velocity \(\mathbf v_t\), and valid domain \(\Omega\). \\
\textbf{Output} & Corrected dynamical state \((A_{t+1}^{D},S_{t+1})\). \\
\textbf{Notation} & \([x]_+=\max(x,0)\), \(\varepsilon>0\) is a stability constant for weight calculation, and \(S_{\max}\) is the prescribed upper bound of the relative ice-mass state. \\
\textbf{Step 1: Baseline transport} & Advance SIC and relative ice mass separately through the existing dynamical transport to obtain the intermediate states \(A^{\mathrm{adv}}\) and \(S^{\mathrm{adv}}\). \\
\textbf{Step 2: Area-deficit calculation} & Because the grid is equal-area, calculate the equivalent sea ice area deficit induced by transport as \(\displaystyle \Delta m=\left[\sum_{p\in\Omega}A_{t,p}-\sum_{p\in\Omega}A_{p}^{\mathrm{adv}}\right]_+\). \\
\textbf{Step 3: Area-constrained redistribution} & Calculate the local loss, remaining capacity, and replenishment weight as \(\displaystyle \ell_p=[A_{t,p}-A_{p}^{\mathrm{adv}}]_+\), \(\displaystyle c_p=[1-A_{p}^{\mathrm{adv}}]_+\), and \(\displaystyle w_p=\ell_pc_p\sqrt{A_{t,p}+\varepsilon}\). Distribute \(\Delta m\) according to the normalized weights \(w_p\), with \(c_p\) limiting local replenishment, to obtain \(A^{\mathrm{area}}\). If all local-loss weights are zero, determine replenishment locations solely from the remaining capacity and existing sea ice distribution. \\
\textbf{Step 4: Compression-overflow and expansion-loss calculation} & Obtain the unclipped state \(A^{\mathrm{u}}\) after the baseline compression and expansion update, and define \(\displaystyle A^{\mathrm{clip}}=\Pi_{[0,1]}(A^{\mathrm{u}})\), \(\displaystyle O=[A^{\mathrm{u}}-1]_+\), and \(\displaystyle L=[A^{\mathrm{area}}-A^{\mathrm{clip}}]_+\). Here, \(O\) denotes compression overflow beyond the SIC upper bound, and \(L\) denotes the explicit SIC loss induced by expansion. \\
\textbf{Step 5: Relative ice-mass update} & Release the recoverable SIC increment from the transported relative ice mass, \(\displaystyle \Delta A^{\mathrm{rel}}=\min(S^{\mathrm{adv}},L)\), and update the coupled dynamical state as \(\displaystyle A_{t+1}^{D}=\Pi_{[0,1]}(A^{\mathrm{clip}}+\Delta A^{\mathrm{rel}})\) and \(\displaystyle S_{t+1}=\Pi_{[0,S_{\max}]}(S^{\mathrm{adv}}+O-\Delta A^{\mathrm{rel}})\). \\
\textbf{Return} & Pass \((A_{t+1}^{D},S_{t+1})\) as the dynamical state for the next time step. \\
\bottomrule
\end{tabular}
\end{table*}

\subsection{Variable-Semantic Adaptation for Forecast Forcing}

Operational forecasting cannot access true forcing fields beyond initialization and must therefore rely on forecast forcing. However, the frozen evolution core represents the mapping between reanalysis states and reanalysis forcing, and direct input replacement would confound forecast-variable errors with base-model simulation errors. Accordingly, the evolution core is kept fixed, with variable-semantic adaptation applied at its input and residual response biases corrected at its output, thereby separating the two error sources. The newly introduced learnable parameters are grouped by function as

\[
\begin{aligned}
\Psi
&=
\left\{
\Psi_{\mathrm{in}},\Psi_{\mathrm{resp}}
\right\},\\
\Psi_{\mathrm{in}}
&=
\left\{
\psi_A,\psi_F
\right\},
\qquad
\psi_F
=
\left\{
\psi_{\mathrm{sca}},\psi_V,\psi_G,\psi_Q
\right\},\\
\Psi_{\mathrm{resp}}
&=
\left\{
\psi_B,\psi_{\mathrm{dyn}},\psi_{\mathrm{frz}},\psi_{\mathrm{mlt}}
\right\}.
\end{aligned}
\tag{14}
\]

Here, \(\Psi\) denotes all parameters introduced in the forecast domain; \(\psi_A\) denotes the observed-state adaptation parameters; \(\psi_F\) denotes the forcing-adaptation parameter set comprising scalar adaptation, vector adaptation, reliability gating, and net-heat-flux adaptation; and \(\Psi_{\mathrm{resp}}\) denotes the response-correction parameters following the frozen evolution core.

\subsubsection{Input-Domain Variable-Semantic Adaptation}

Before entering the frozen evolution core, the observational initial state and forecast forcing undergo deterministic alignment of grids, units, sign conventions, and vector orientations. Because observed SIC, scalar forcing, and vector sea ice velocity exhibit state biases, statistical distribution shifts, and magnitude--direction errors, respectively, each variable is adapted according to its physical semantics. The objective is not to reconstruct unknown true fields, but to construct equivalent inputs compatible with the reanalysis-domain evolution core.

For the observational initial state, the model estimates a local residual conditioned on SIC, the ice-edge gradient, and environmental context:

\[
\begin{aligned}
 A_t^{c}
 &=
 \mathcal A_{\psi_A}^{A}\left(A_t^{\mathrm{obs}}\right)\\
 &=
 \Pi_{[0,1]}
 \left[
 A_t^{\mathrm{obs}}
 +\Delta A_{\psi_A}
 \left(
 A_t^{\mathrm{obs}},\nabla A_t^{\mathrm{obs}},\mathbf c_t
 \right)
 \right].
\end{aligned}
\tag{15}
\]

Here, \(A_t^c\) denotes the calibrated initial state, and \(\mathbf c_t\) represents seasonal, spatial-location, and data-coverage information. This procedure corrects only local discrepancies between the observational product and the reanalysis state, without regenerating the initial sea ice field.

For scalar forcing variables, including sea surface temperature, 2-m air temperature, and net heat flux, the model jointly corrects lead-dependent scaling, systematic bias, and local spatial residuals:

\[
\overline z_{t+k-1}^{m}
=
\left(1+s_k^{m}\right)z_{t+k-1}^{m}
+
b_k^{m}
+
r_{\psi_{\mathrm{sca}}}^{m}
\left(z_{t+k-1}^{m},\mathbf c_{t+k-1}\right).
\tag{16}
\]

Here, \(t\) is the initialization time, \(k\) is the relative time step of the target day, and \(m\) indexes the forcing variable; \(z_{t+k-1}^{m}\) is the forecast field that drives evolution on day \(t+k\) after deterministic alignment and standardization against reanalysis statistics. The scalar-adaptation parameters \(\psi_{\mathrm{sca}}\) comprise the learnable scale \(s_k^m\), bias \(b_k^m\), and local spatial residual operator \(r_{\psi_{\mathrm{sca}}}^m\). The atmospheric adapter jointly processes 2-m air temperature and net heat flux, whereas an independent ocean adapter processes sea surface temperature.

Sea ice velocity comprises two correlated vector components; its speed scale, direction, and local residual are therefore corrected jointly:

\[
\overline{\mathbf v}_{t+k-1}
=
s_k^{v}\mathbf R(\alpha_k)
\mathbf v_{t+k-1}^{\mathrm{fcst}}
+
\Delta\mathbf v_{\psi_V}
\left(
\mathbf v_{t+k-1}^{\mathrm{fcst}},
\mathbf c_{t+k-1}
\right).
\tag{17}
\]

Here, the sea ice velocity adaptation parameters \(\psi_V\) comprise the lead-dependent speed scale \(s_k^v\), rotation correction \(\alpha_k\), and local vector residual operator \(\Delta\mathbf v_{\psi_V}\), while \(\mathbf R(\alpha_k)\) denotes the corresponding rotation operator. This design corrects forecast errors while preserving the vector relationship between the two sea ice velocity components.

These four module groups jointly parameterize the forcing adaptation operator \(\mathcal A_{\psi_F}^{F}\) in Eq. (4). The sea ice velocity adapter \(\mathcal V_{\psi_V}\) combines velocity structure and ice-edge state to produce a local vector residual and separate gates for the physical baseline and compensation network. Following general scalar calibration, the net-heat-flux adapter \(\mathcal M_{\psi_Q}\) further draws on flux, temperature, and initial-state information within reliable forecast leads to learn a flux-specific residual, prefix thermodynamic memory, and dual-path gating; the pathway weights for sea surface temperature and 2-m air temperature are determined by the season--lead gate \(g_{\psi_G}^{p,m}\), where \(p\in\{\mathcal P,\mathcal C\}\) denotes the physical baseline or compensation network. Gradients, velocity diagnostics, and temporal-prefix accumulation are deterministic features; only the residual mappings and gating parameters are learnable.

The shared local residual unit and the layer-wise architectures, output constraints, and parameter counts of the input adapters are provided in Supplementary Materials E.1 and E.2.

\subsubsection{Response-Domain Dynamical--Thermodynamic Semantic Correction}

Input-domain adaptation can only reduce semantic discrepancies between forecast and reanalysis variables; it cannot recover unknown true forcing. Let \(A_{t+k}^{\mathrm{core}}\) denote the \(k\)-th-step output produced by the frozen evolution core after input adaptation. After nonlinear propagation through the frozen core, the remaining errors further manifest as SIC response biases. Accordingly, a learnable base semantic response head \(\mathcal B_{\psi_B}\) first maps the core output to the observation domain:

\[
A_{t+k}^{\mathrm{base}}
=
\mathcal B_{\psi_B}
\left(
A_{t+k}^{\mathrm{core}},
\mathbf d_{t+k}^{B}
\right),
\tag{18}
\]

Here, \(\mathbf d_{t+k}^{B}\) comprises the physical trajectory, dynamical--thermodynamic diagnostics, and environmental context. The remaining biases are further decomposed into dynamical, freezing, and melting responses and corrected separately.

The dynamical branch does not repeat sea ice transport; instead, it constructs dynamical response diagnostics from the effect of adapted sea ice velocity on the physical trajectory. Setting \(\widetilde A_t=A_t^c\) forms a continuous state sequence from the calibrated initial state and subsequent physical trajectory; the local advective tendency and its prefix-cumulative response are then expressed as

\[
\begin{aligned}
\tau_{t+k}^{D}
&=
-\overline{\mathbf v}_{t+k-1}
\cdot
\nabla \widetilde A_{t+k-1},\\
M_{t+k}^{D}
&=
\frac{1}{\sqrt{k}}
\sum_{j=1}^{k}
\tau_{t+j}^{D}.
\end{aligned}
\tag{19}
\]

Here, \(\tau_{t+k}^{D}\) characterizes the instantaneous effect of adapted sea ice velocity on the local sea ice distribution, while \(M_{t+k}^{D}\) represents the cumulative influence of preceding dynamical effects. Together with velocity divergence, vorticity, and the ice-edge gradient, these quantities constitute the dynamical diagnostics \(\mathbf d_{t+k}^{D}\), from which the learnable dynamical response head produces

\[
\Delta A_{t+k}^{D}
=
\mathcal H_{\psi_{\mathrm{dyn}}}^{D}
\left(
\mathbf d_{t+k}^{D}
\right).
\tag{20}
\]

These diagnostics do not recompute the physical trajectory and serve only as conditioning information for the dynamical response head.

The thermodynamic response is divided into freezing and melting branches according to phase-change direction. Let \(\overline Q_{t+j-1}\) denote the adapted and standardized net heat flux driving evolution on day \(t+j\), with positive and negative values indicating melting and freezing tendencies, respectively. The model constructs prefix heat-flux memories over the prefix available at the current time step:

\[
\begin{aligned}
 M_{t+k}^{\mathrm{freeze}}
 &=
 \frac{1}{\sqrt{k}}
 \sum_{j=1}^{\min(k,K_Q)}[-\overline Q_{t+j-1}]_+,\\
 M_{t+k}^{\mathrm{melt}}
 &=
 \frac{1}{\sqrt{k}}
 \sum_{j=1}^{\min(k,K_Q)}[\overline Q_{t+j-1}]_+.
\end{aligned}
\tag{21}
\]

Here, \([x]_+=\max(x,0)\), and \(K_Q\) denotes the reliable forecast range of net heat flux. The factor \(1/\sqrt{k}\) suppresses persistent growth of the accumulated features with lead time and reduces the influence of historical flux beyond the reliable forecast range.

Freezing-flux memory, low-temperature conditions, and open-water capacity jointly form the freezing diagnostics \(\mathbf d_{t+k}^{\mathrm{frz}}\); melting-flux memory, high-temperature conditions, and existing sea ice jointly form the melting diagnostics \(\mathbf d_{t+k}^{\mathrm{mlt}}\). Two response heads with identical architectures but independent parameters respectively produce

\[
\Delta A_{t+k}^{\mathrm{freeze}}
=
\mathcal H_{\psi_{\mathrm{frz}}}^{\mathrm{freeze}}
\left(
\mathbf d_{t+k}^{\mathrm{frz}}
\right),
\qquad
\Delta A_{t+k}^{\mathrm{melt}}
=
\mathcal H_{\psi_{\mathrm{mlt}}}^{\mathrm{melt}}
\left(
\mathbf d_{t+k}^{\mathrm{mlt}}
\right).
\tag{22}
\]

Here, 2-m air temperature and sea surface temperature provide the phase-change conditions, while the current sea ice state constrains the spatial support of the corresponding processes; state-support and sign constraints preserve the principal directions of freezing and melting corrections.

The final prediction combines the base response with the three semantic corrections:

\[
\widehat A_{t+k}^{\mathrm{fcst}}
=
\Pi_{[0,1]}
\left[
A_{t+k}^{\mathrm{base}}
+
\Delta A_{t+k}^{D}
+
\Delta A_{t+k}^{\mathrm{freeze}}
+
\Delta A_{t+k}^{\mathrm{melt}}
\right].
\tag{23}
\]

The base semantic response head and the three specialized response heads are lightweight residual networks whose parameters jointly constitute \(\Psi_{\mathrm{resp}}\). The three specialized responses are estimated independently and combined additively, without explicit dynamical--thermodynamic interaction parameters. All diagnostics and corrections depend only on information from the current and preceding time steps and do not alter the sea ice evolution relationships learned by the frozen evolution core. The architectures and parameter counts of the response heads are provided in Supplementary Materials E.3.

\subsection{Stage-Wise Optimization for Dual-Domain Decoupling}

To prevent forecast-input errors from altering the sea ice evolution relationships learned in the reanalysis domain, a progressive-freezing optimization strategy sequentially learns the physical-baseline parameters, joint multi-lead compensation parameters, and forecast-domain adaptation parameters. Each stage updates only its designated parameter subset, and all spatial losses are evaluated within the valid-ocean mask.

\subsubsection{Optimization of the Reanalysis-Domain Evolution Core}

The thermodynamic parameters \(\theta\) of the physical baseline are first optimized on reanalysis data, forming stable physical trajectories together with deterministic dynamical transport. The physical baseline is then frozen, and only the joint multi-lead compensation parameters \(\phi\) are optimized, enabling the compensation network to represent unresolved processes and residual errors of the physical baseline. To preserve the daily evolution relationship and input semantics defined by Eq. (1) during large-scale GPU-batched training, a lightweight physical baseline with explicit process semantics first generates and caches multi-lead physical trajectories through daily propagation, thereby carrying state recurrence; Eq. (5) then organizes each propagated state and its corresponding forcing as compensation input, enabling the compensation network to jointly estimate multi-lead residuals in a single forward pass without participating in serial state propagation, thereby improving GPU utilization and supporting greater representational capacity. The objective function for the compensation stage is

\[
\mathcal L_{\mathrm{core}}
=
\mathcal L_{\mathrm{SIC}}
+\lambda_b\mathcal L_{\mathrm{bias}}
+\lambda_r\mathcal L_{\mathrm{res}}.
\tag{24}
\]

Here, \(\mathcal L_{\mathrm{SIC}}\) is the multi-lead SIC mean squared error within the valid domain, \(\mathcal L_{\mathrm{bias}}\) is the squared difference between the regional means of the prediction and target, and \(\mathcal L_{\mathrm{res}}\) is the \(L_1\) regularization term on the compensation residuals. The explicit forms and weights of these loss terms are provided in Supplementary Materials A.4. Let \(\Theta=(\theta,\phi)\) denote the evolution-core parameter set. Training yields

\[
\Theta^*=(\theta^*,\phi^*).
\tag{25}
\]

The physical baseline and joint multi-lead compensation network jointly constitute the reanalysis-domain evolution core and remain frozen during subsequent forecast-domain adaptation.

\subsubsection{Optimization of Forecast-Domain Adaptation}

Throughout forecast-domain optimization, \(\Theta=\Theta^*\) remains fixed, and only the newly introduced adaptation parameters are learned. The sea ice velocity adaptation parameters are further decomposed as \(\psi_V=\{\psi_V^{\mathrm{base}},\psi_V^{\mathrm{trust}}\}\), denoting the base velocity-calibration parameters and the velocity-reliability diagnostic and gating parameters, respectively. The parameters introduced for forecast-domain adaptation are grouped by function as

\[
\begin{aligned}
\Psi_1
&=
\left\{
\psi_A,
\psi_{\mathrm{sca}},
\psi_V^{\mathrm{base}}
\right\},\\
\Psi_2
&=
\left\{
\psi_V^{\mathrm{trust}},
\psi_G,
\psi_Q,
\psi_B
\right\},\\
\Psi_3
&=
\left\{
\psi_{\mathrm{dyn}},
\psi_{\mathrm{frz}},
\psi_{\mathrm{mlt}}
\right\}.
\end{aligned}
\tag{26}
\]

The first stage optimizes the base input-adaptation parameters \(\Psi_1\), establishing a base mapping from observational initial states and forecast forcing to reanalysis-domain inputs through observation--reanalysis state alignment, forecast--reanalysis forcing alignment, and evolution-response matching under the frozen core. The second stage freezes the base adapters and optimizes the variable-semantic adaptation and base-response parameters \(\Psi_2\); observed SIC constrains the predictions, while variable anchoring and magnitude regularization limit overcorrection. The third stage freezes all preceding parameters and optimizes only the specialized response parameters \(\Psi_3\), further correcting the dynamical, freezing, and melting responses.

The three stages are uniformly formulated as

\[
\begin{aligned}
\Psi_s^*&=\underset{\Psi_s}{\arg\min}\,\mathcal L_s,\\
\mathrm{s.t.}\quad
&\Theta=\Theta^*,\qquad
\Psi_{<s}=\Psi_{<s}^*,\qquad s=1,2,3.
\end{aligned}
\tag{27}
\]

Here, \(\Psi_{<s}\) denotes the parameter sets optimized before stage \(s\). The objective \(\mathcal L_1\) comprises state and forcing alignment, evolution-response matching, and correction-magnitude terms; \(\mathcal L_2\) comprises terms for SIC, sea ice area, sea ice extent, ice-edge structure, variable anchoring, and correction magnitude; and \(\mathcal L_3\) further incorporates freezing--melting directional constraints, a reanalysis-domain identity constraint, and branch regularization in addition to observation-based constraints. The explicit forms and weights of the stage-wise losses are provided in Supplementary Materials A.4, and the parameter subsets and network architectures are provided in Supplementary Materials E.1--E.4.

During forecast-domain adaptation, contemporaneous reanalysis data contribute only to state and forcing alignment, evolution-response matching, and identity-preservation constraints; operational forecasting relies solely on the observational initial state and forecast forcing, without future reanalysis fields or future observations.

\section{Experiments}

The experiments proceed from validation of the base evolution core to adaptation of forecast-domain errors. First, the model's capability to simulate sea ice evolution under reanalysis initial states and forcing is evaluated, while module-ablation and variable-sensitivity experiments examine its underlying mechanisms. The base evolution core is subsequently frozen, and the model's forecast performance after variable-semantic adaptation is assessed under observational initial states and forecast forcing. All experiment groups follow standardized and mutually independent evaluation protocols.

\subsection{Datasets and Experimental Settings}

The experiments comprise two tasks: reanalysis-forced simulation and forecast-forced prediction. Both tasks target daily SIC on a \(432\times432\) Arctic EASE-Grid 2.0 (EASE2) equal-area grid with a spatial resolution of 25 km \cite{ref51,ref52}. Table~\ref{tab:4-1} summarizes the data sources, temporal splits, and evaluation lead times.

\begin{table*}[!t]
\centering
\scriptsize
\caption{Data Sources, Temporal Splits, and Evaluation Protocols.}
\label{tab:4-1}
\begin{tabular}{@{}
  >{\raggedright\arraybackslash}p{(\textwidth - 12\tabcolsep) * \real{0.15}}
  >{\raggedright\arraybackslash}p{(\textwidth - 12\tabcolsep) * \real{0.11}}
  >{\raggedright\arraybackslash}p{(\textwidth - 12\tabcolsep) * \real{0.30}}
  >{\raggedright\arraybackslash}p{(\textwidth - 12\tabcolsep) * \real{0.10}}
  >{\raggedright\arraybackslash}p{(\textwidth - 12\tabcolsep) * \real{0.10}}
  >{\raggedleft\arraybackslash}p{(\textwidth - 12\tabcolsep) * \real{0.14}}
  >{\raggedleft\arraybackslash}p{(\textwidth - 12\tabcolsep) * \real{0.10}}@{}}
\toprule
Task & Initial State and Target & External Forcing & Training or Adaptation Period & Evaluation Period & No.~of Initializations (Training/Evaluation) & Lead Time \\
\midrule
Reanalysis-forced simulation & GLORYS12V1 SIC \cite{ref53} & GLORYS12V1 sea ice velocity and sea surface temperature \cite{ref53}; ERA5 2-m air temperature and net heat flux \cite{ref54} & 2015--2021 & 2022--2024 & 2459/1054 & 14 days \\
Forecast-forced prediction & OSI-430-a and OSI-401-d SIC \cite{ref55,ref56} & TOPAZ5 sea ice velocity and sea surface temperature \cite{ref57}; GFS 2-m air temperature and net heat flux \cite{ref58} & 2023-09-01--2024-08-31 & 2024-09-18--2025-09-01 & 366/336 & 10 days \\
\bottomrule
\end{tabular}
\end{table*}

During forecast-forced evaluation, only observed initial SIC and forecast forcing are provided; neither future reanalysis fields nor future observations are accessed.

The evaluation metrics comprise mean absolute error (MAE), root mean squared error (RMSE), mean bias (Bias), anomaly correlation coefficient (ACC), integrated ice-edge error (IIEE), and absolute sea ice area error (SIAE). IIEE adopts a 15\% SIC threshold \cite{ref59}. Data processing and sample selection, metric definitions, and the comparison models and training configurations are detailed in Supplementary Materials A.1, A.2, and A.3, respectively.

\subsection{Reanalysis-Forced Simulation of Sea Ice Evolution}

This section first examines whether the base evolution core can accurately simulate daily sea ice evolution under high-quality reanalysis initial states and forcing. The proposed model is compared with representative spatiotemporal models in terms of overall, lead-wise, and ice-edge errors, followed by annual closed-loop simulation to evaluate long-term evolution stability.

\subsubsection{Overall and Lead-Wise Simulation Performance}

Table~\ref{tab:4-2} reports the 14-day mean performance of all models. Overall, the proposed model exhibits consistent advantages in grid-point error, anomaly correlation, and ice-edge position while maintaining a low systematic bias, indicating improvements in both overall accuracy and spatial distribution.

\begin{table}[!t]
\centering
\scriptsize
\caption{Fourteen-Day Mean Performance Under Reanalysis Forcing.}
\label{tab:4-2}
\resizebox{\columnwidth}{!}{%
\begin{tabular}{@{}lrrrrr@{}}
\toprule
Model & MAE & RMSE & Bias & ACC & IIEE \\
\midrule
PredRNN++ & 0.031302 & 0.083033 & +0.002932 & 0.808657 & 0.458069 \\
FCNet & 0.029835 & 0.081050 & \textbf{+0.000717} & 0.814477 & 0.404746 \\
VMRNN & 0.029337 & 0.071583 & +0.002563 & 0.855370 & 0.369199 \\
SwinLSTM & 0.029252 & 0.071281 & +0.003275 & 0.856853 & 0.362219 \\
PredRNN-v2 & 0.026686 & 0.068568 & +0.000905 & 0.867448 & 0.361438 \\
SimVP & 0.024916 & 0.067109 & +0.001198 & 0.874807 & 0.284190 \\
IceNet & 0.023978 & 0.065669 & +0.003089 & 0.882835 & 0.270330 \\
Ours & \textbf{0.020235} & \textbf{0.052285} & +0.002911 & \textbf{0.925123} & \textbf{0.229394} \\
\bottomrule
\end{tabular}
}
\end{table}

\begin{figure}[!t]
\centering
\includegraphics[width=0.8\columnwidth]{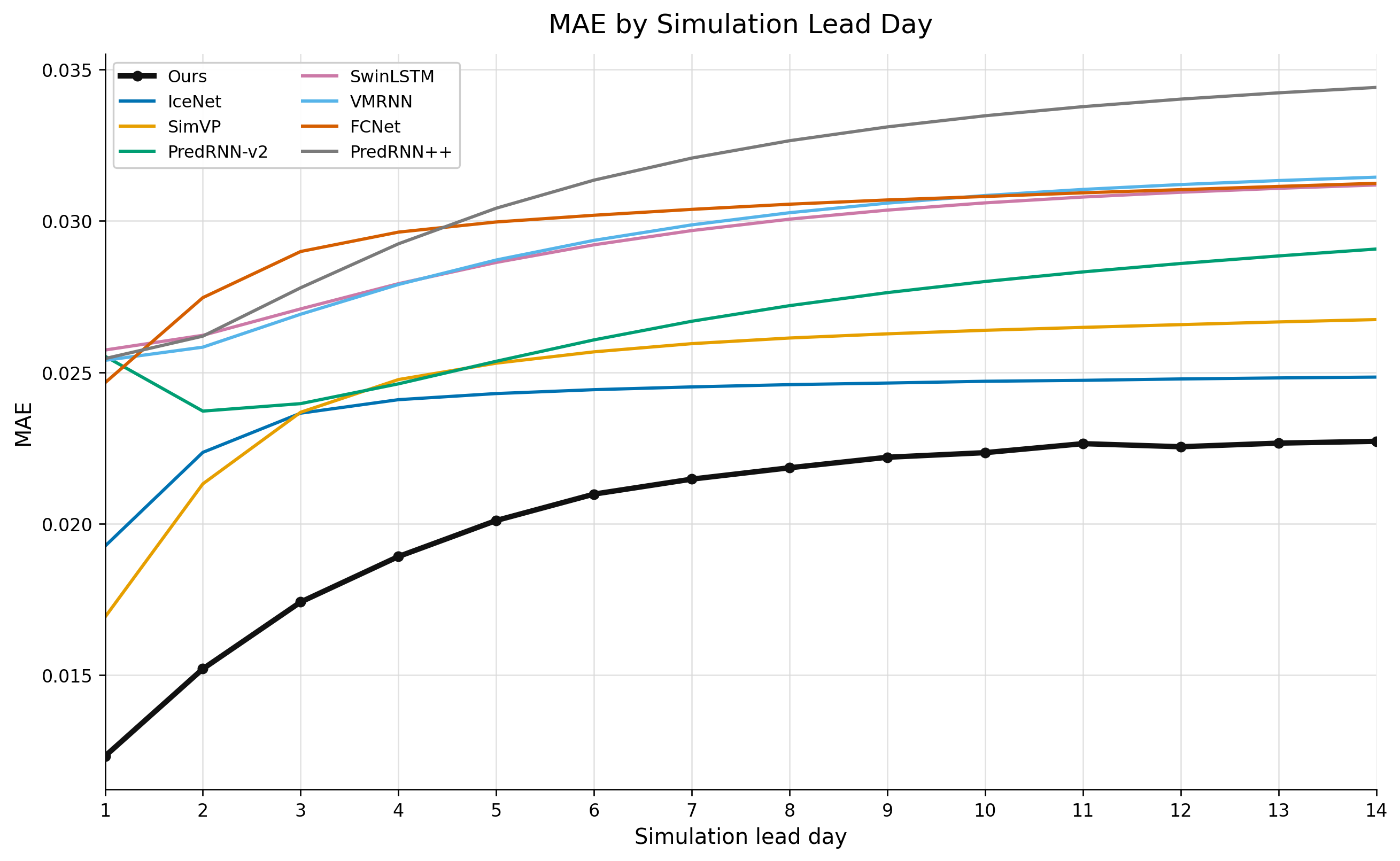}
\caption{SIC MAE of all models over days 1--14.}
\label{fig:4-1}
\end{figure}

\begin{figure}[!t]
\centering
\includegraphics[width=0.8\columnwidth]{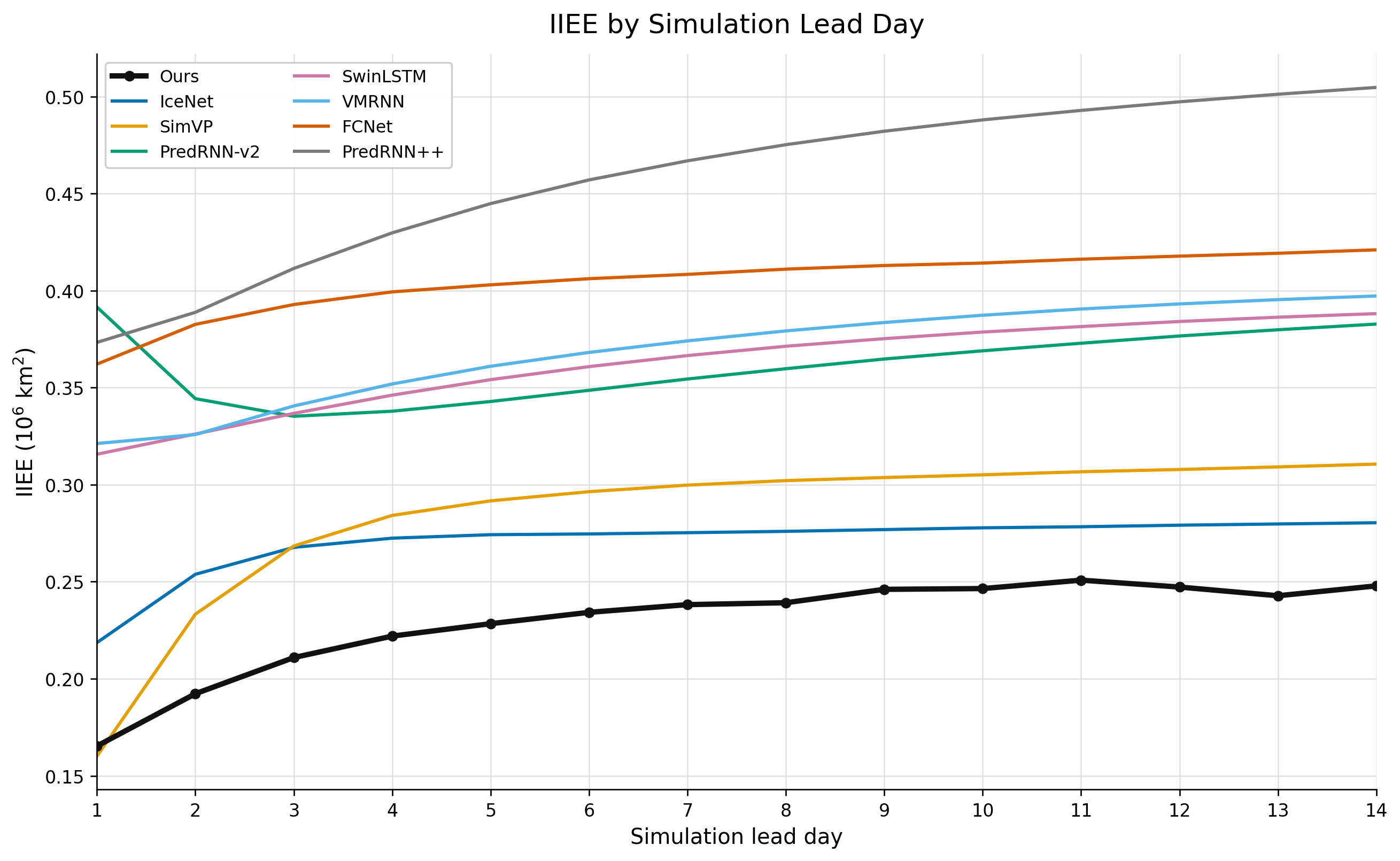}
\caption{SIC IIEE of all models over days 1--14.}
\label{fig:4-2}
\end{figure}

Fig.~\ref{fig:4-1} and Fig.~\ref{fig:4-2} show the lead-dependent MAE and IIEE of all models. The proposed model maintains the lowest MAE at every lead and the lowest IIEE from day 2 onward. Both errors tend to stabilize at later leads, indicating that the advantages in overall accuracy and ice-edge simulation persist throughout continuous evolution.

\subsubsection{Long-Term Evolution Stability}

\begin{figure}[!t]
\centering
\includegraphics[width=0.8\columnwidth]{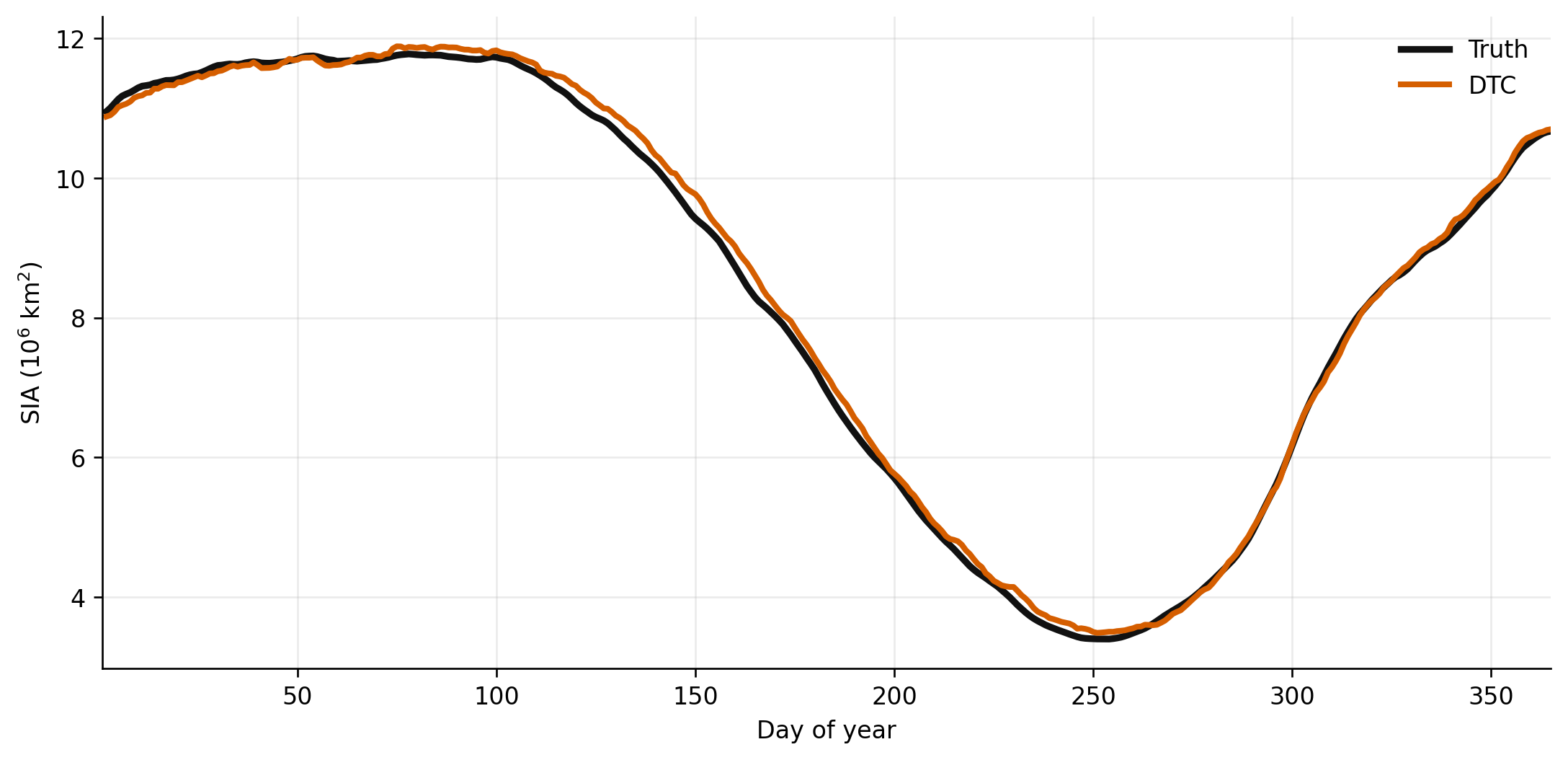}
\caption{Mean annual SIA cycle over 2022--2024.}
\label{fig:4-3}
\end{figure}

To assess model stability over timescales far exceeding the 14-day output horizon, annual closed-loop simulations were conducted separately for 2022--2024. Reanalysis SIC was supplied only once on 1 January of each year; thereafter, the final-day output of each 14-day simulation block served as the initial condition for the subsequent block, with no further reanalysis SIC introduced. Fig.~\ref{fig:4-3} shows the three-year mean annual cycle of sea ice area (SIA). This section considers only the complete model and the reanalysis reference, while the decomposition of the physical baseline and compensation module is examined in Section IV-C.

The complete model reproduces the freezing, melting, and refreezing phases. Across the three annual simulations, the mean SIA correlation coefficient and seasonal amplitude ratio are 0.99914 and 0.99530, respectively; the mean end-of-year SIA bias is \(+0.027\times10^6\ \mathrm{km}^2\), and the pooled annual SIC MAE is 0.02412. This MAE is derived from three annual closed-loop integrations and therefore follows a different evaluation protocol from the independent 14-day samples in Table~\ref{tab:4-2}. These results indicate that repeated cross-block transfer of the SIC state does not cause apparent divergence and that the model stably preserves the phase and amplitude of the seasonal sea ice cycle.

\subsection{Analysis of Model Mechanisms and Contributions}

Section IV-B demonstrated the model's simulation capability under reanalysis forcing. This section examines the functional division and ice-mass conservation of the physical baseline and compensation network, verifies the temporal constraints and output-prefix invariance of the joint multi-lead compensation network, and quantifies the pathway-specific contributions of environmental variables, providing a reference for the forecast-domain adaptation examined in Section IV-D.

\subsubsection{Functional Division between the Physical Baseline and Conditional Residual Compensation}

\textbf{Stagewise States and Functional Division}

To examine the role of each module within the complete model, the persistence state (Persistence), dynamical output (D), physical-baseline output (DT), and final compensated output (DTC) are sequentially extracted along the forward computation pathway from the same test samples. Here, D represents dynamical transport, DT incorporates thermodynamic source and sink terms into D, and DTC further adds the compensation residual to the DT physical trajectory. Table~\ref{tab:4-3} and Table~\ref{tab:4-4} report the 14-day mean and annual closed-loop results for these states.

\begin{table}[!t]
\centering
\scriptsize
\caption{Fourteen-Day Mean Performance of the Stagewise States.}
\label{tab:4-3}
\resizebox{\columnwidth}{!}{%
\begin{tabular}{@{}lrrrrr@{}}
\toprule
State & MAE & RMSE & Bias & ACC & IIEE \\
\midrule
Persistence & 0.038852 & 0.109363 & +0.000894 & 0.699885 & 0.620202 \\
D & 0.043211 & 0.115129 & +0.000110 & 0.673471 & 0.612158 \\
DT & 0.036729 & 0.093969 & -0.001214 & 0.738146 & 0.528937 \\
DTC & \textbf{0.020235} & \textbf{0.052285} & +0.002911 & \textbf{0.925123} & \textbf{0.229394} \\
\bottomrule
\end{tabular}
}
\end{table}

\begin{table}[!t]
\centering
\scriptsize
\caption{Annual Closed-Loop Performance of the Stagewise States.}
\label{tab:4-4}
\resizebox{\columnwidth}{!}{%
\begin{tabular}{@{}lrrr@{}}
\toprule
State & Annual SIC MAE & SIA Seasonal Correlation & SIA Amplitude Ratio \\
\midrule
D & 0.189529 & 0.643167 & 0.145488 \\
DT & 0.058713 & 0.992654 & 1.144676 \\
DTC & \textbf{0.024120} & \textbf{0.999362} & \textbf{1.002461} \\
\bottomrule
\end{tabular}
}
\end{table}

The annual SIC MAE is pooled across the three annual integrations, whereas the SIA seasonal correlation and amplitude ratio are computed from the three-year mean annual cycle.

\begin{figure}[!t]
\centering
\includegraphics[width=\columnwidth]{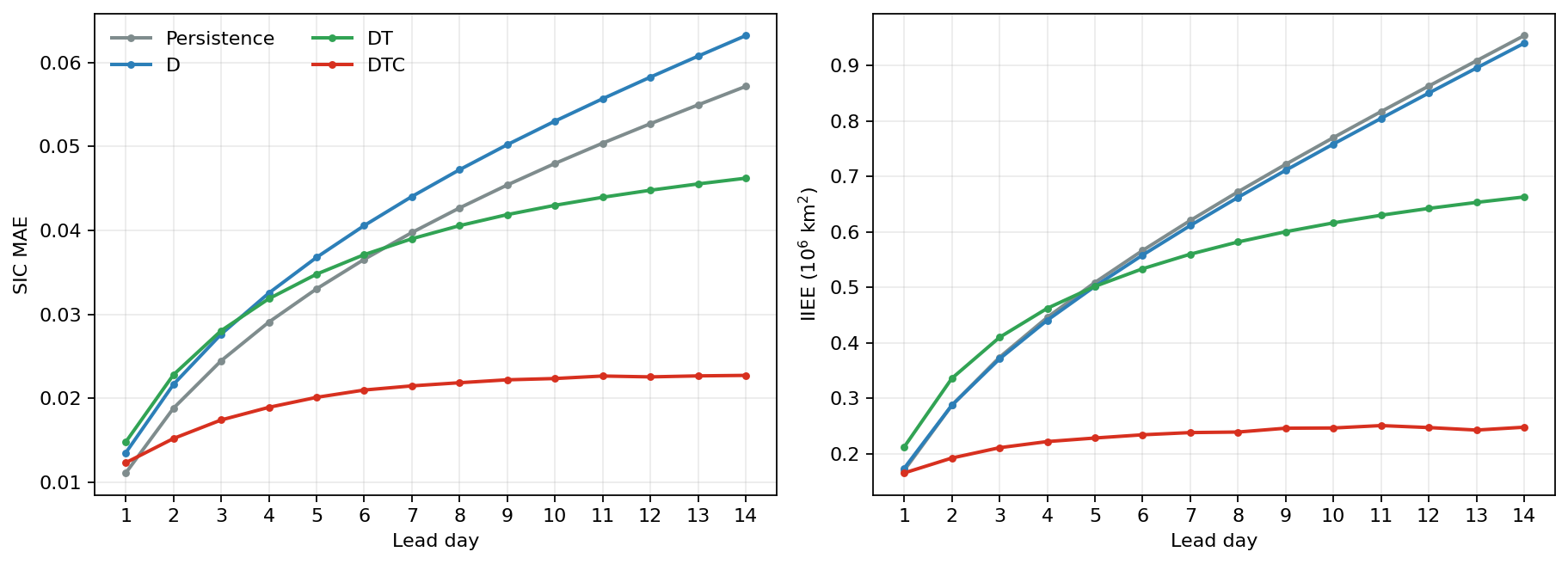}
\caption{Lead-wise errors of Persistence, D, DT, and DTC over 14 days.}
\label{fig:4-4}
\end{figure}

As shown in Table~\ref{tab:4-3} and Fig.~\ref{fig:4-4}, D increases MAE by 11.2\% relative to Persistence but reduces IIEE by 1.3\% and decreases the absolute Bias, indicating that the dynamics module primarily performs spatial transport rather than independently representing changes in total ice amount. After thermodynamic source and sink terms are incorporated, the MAE of DT decreases by 15.0\% relative to D. This contribution is limited during the first three lead times but becomes consistently beneficial from day 4 onward, indicating that freezing and melting effects accumulate with integration time. The compensation module further reduces the MAE of DT across all 14 lead times, yielding an overall reduction of 44.9\%.

\begin{figure}[!t]
\centering
\includegraphics[width=0.8\columnwidth]{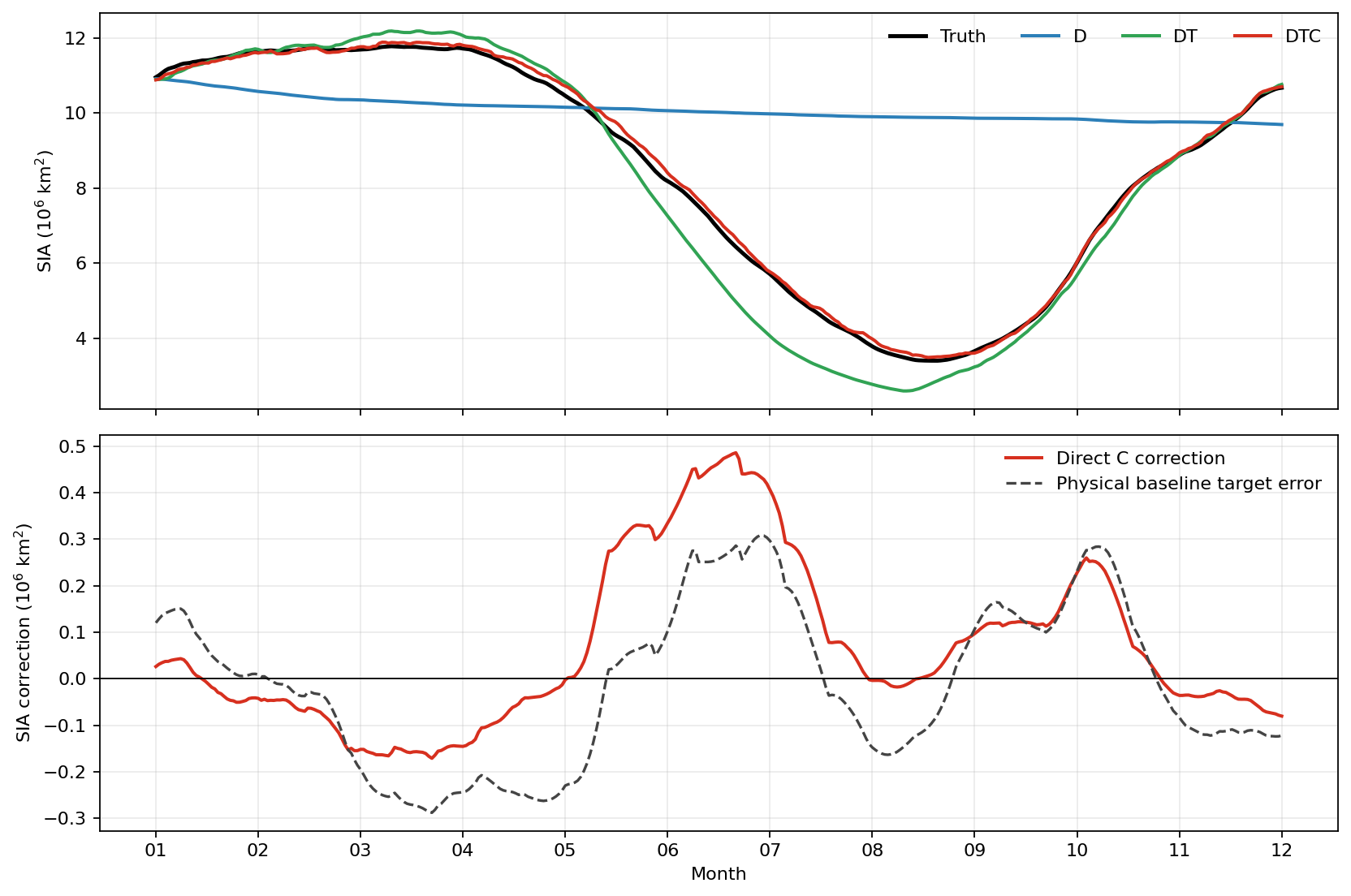}
\caption{Three-year mean annual SIA evolution over 2022--2024 and direct correction of physical-baseline errors by the compensation module.}
\label{fig:4-5}
\end{figure}

The corresponding annual SIA trajectories are shown in Fig.~\ref{fig:4-5}.

As shown in Table~\ref{tab:4-4} and Fig.~\ref{fig:4-5}, the annual closed-loop results further confirm this functional division. D alone cannot maintain a physically reasonable freezing--melting cycle. DT recovers the seasonal cycle but overestimates its amplitude, whereas DTC reduces the annual SIC MAE by 58.9\% relative to DT and brings both the correlation coefficient and amplitude ratio of the three-year mean SIA cycle close to the ideal value of 1. Event-scale responses and initialization-offset sensitivity are provided in Supplementary Materials B.1.

\begin{figure}[!t]
\centering
\includegraphics[width=\columnwidth]{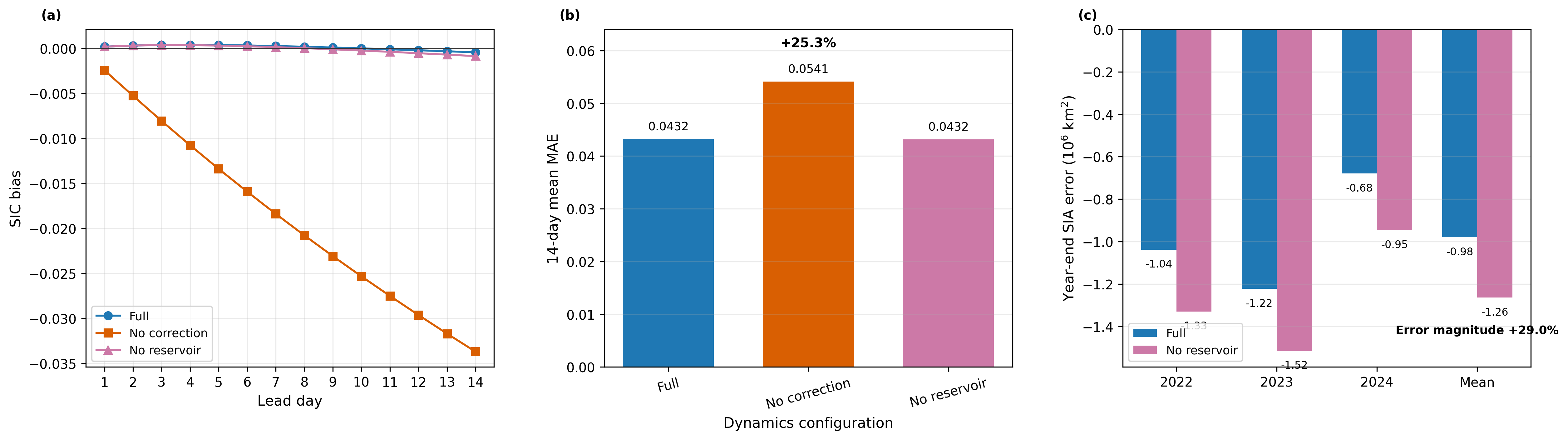}
\caption{Ablation diagnostics of the ice-mass conservation mechanisms: (a) lead-wise SIC Bias for the complete dynamics module, without mass refilling, and without relative ice-mass storage; (b) 14-day mean MAE under the three dynamics configurations; (c) annual and three-year mean end-of-year SIA errors for the complete dynamics module and the configuration without relative ice-mass storage.}
\label{fig:4-6}
\end{figure}

\textbf{Ice-Mass Conservation in the Dynamics Module.} Because D primarily performs spatial transport, its ice-mass conservation during recursive integration must also be examined. Fig.~\ref{fig:4-6} presents ablation results for lead-wise Bias, 14-day mean MAE, and end-of-year SIA error.

As shown in Fig.~\ref{fig:4-6}(a) and (b), removing mass refilling increases the 14-day mean MAE by 25.3\% and substantially enlarges the negative bias on day 14, indicating that mass refilling primarily suppresses short-term nonconservative loss. As shown in Fig.~\ref{fig:4-6}(c), removing relative ice-mass storage has little effect on short-term MAE but increases the magnitude of the three-year mean end-of-year negative SIA bias by 29.0\%, indicating that this mechanism primarily improves long-term ice-mass conservation. Complete results for the individual mechanisms are provided in Supplementary Materials B.2. The annual dynamics-only experiment serves solely to diagnose cumulative ice-mass drift and does not imply that the dynamics module alone can simulate a full annual sea ice cycle.

\begin{figure}[!t]
\centering
\includegraphics[width=\columnwidth]{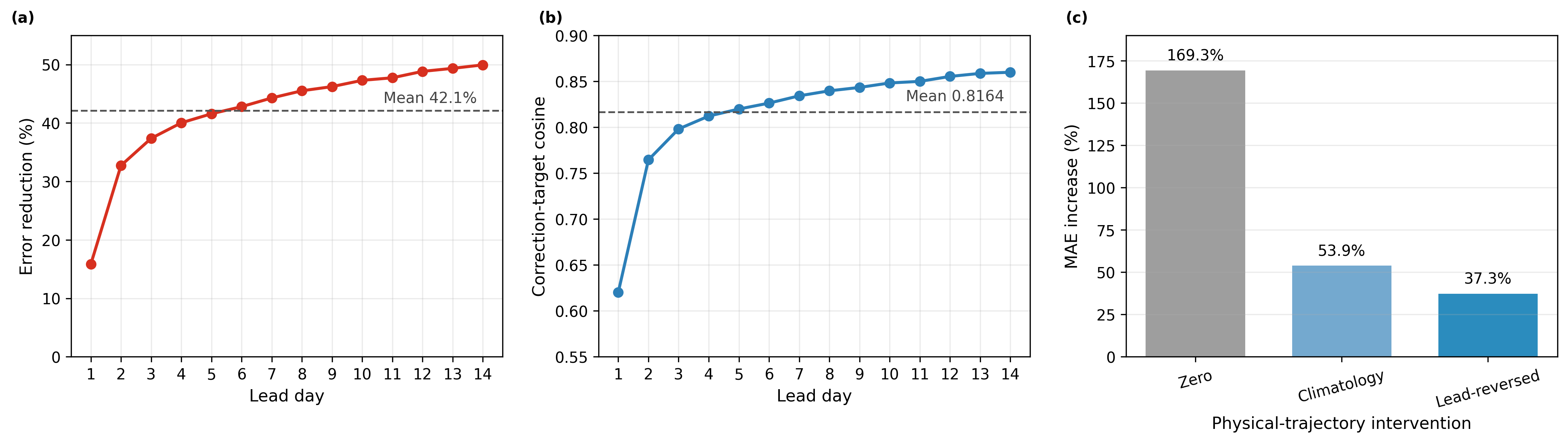}
\caption{Conditional-residual semantics and physical-trajectory dependence of the compensation module: (a) lead-wise error-reduction rate; (b) spatial cosine similarity between the compensation correction and the target physical-baseline error; (c) MAE increases induced by zero, climatological, and lead-reversed trajectories relative to the correct physical trajectory.}
\label{fig:4-7}
\end{figure}

\textbf{Conditional Residual Semantics and Dependence on Physical Trajectories.} To determine whether compensation module C preserves its conditional-residual semantics, the time and lead indices are omitted, the remaining physical-baseline error is denoted by \(E^{\mathrm{DT}}=A^{\mathrm{rean}}-A^{\mathrm{DT}}\), and the actual state correction generated by C is denoted by \(\Delta A^C=A^{\mathrm{DTC}}-A^{\mathrm{DT}}\). The corresponding diagnostics are presented in Fig.~\ref{fig:4-7}.

As shown in Fig.~\ref{fig:4-7}(a) and (b), pointwise state corrections yield a mean error-reduction rate of 42.11\% for C and a spatial cosine similarity of 0.8164 with the target physical-baseline error, indicating that the compensation predominantly acts toward correcting the remaining baseline error. As shown in Fig.~\ref{fig:4-7}(c), when the DT state receiving the final correction is retained but the correct physical trajectory supplied to C is replaced by a zero, climatological, or lead-reversed trajectory, MAE increases by 169.3\%, 53.9\%, and 37.3\%, respectively. This result demonstrates that the compensation module explicitly depends on the physical states and their lead-time correspondence, rather than treating the physical trajectory as an arbitrarily replaceable auxiliary input. Overall, the complete model establishes a progressive functional division comprising dynamical transport, thermodynamic source and sink terms, and conditional residual compensation.

Supplementary comparisons after independent retraining of each architecture are provided in Supplementary Materials B.4.

\subsubsection{Temporal Constraints and Output-Prefix Invariance of Joint Multi-Lead Outputs}

The compensation network jointly generates residuals for 14 lead times in a single forward pass. To satisfy the lead-prefix constraint of daily evolution, the output at lead \(k\) may depend only on the physical states and forcing fields from leads 1 through \(k\), and must remain unaffected by inputs from leads \(k+1\) through 14. Accordingly, the existing output prefix should remain invariant under arbitrary replacement of subsequent inputs. This constraint is implemented through the lower-triangular temporal mixing described in Section III-B2.

\textbf{Temporal-Constraint Structure Verification.} To distinguish the structural effect of the temporal constraint from that of network capacity, three configurations are compared: H24 without temporal constraints, H24 with temporal constraints, and H34 with temporal constraints, where H24 and H34 denote hidden widths of 24 and 34, respectively. H24 without temporal constraints permits unrestricted cross-temporal feature mixing and serves as the unconstrained reference. H24 with temporal constraints retains the same hidden width and parameter count, thereby isolating the effect of the temporal constraint itself. H34 with temporal constraints increases network capacity while preserving the temporal-constraint structure, thereby assessing whether enhanced representation capacity can mitigate the accuracy loss induced by the temporal constraint.

Across all 1,054 test samples, the inputs for days 1--7 are held fixed, while the inputs for days 8--14 are separately replaced by zero forcing, noisy forcing, all-zero inputs, repeated day-7 inputs, and reversed tail inputs. Table~\ref{tab:4-5} reports the mean absolute changes in the original compensation residuals over the first seven days under the different replacement schemes. This metric is defined as

\[
D_{\mathrm{pre}}
=
\frac{1}{7N|\Omega|}
\sum_{\substack{1\le n\le N,\;1\le k\le 7\\ \mathbf x\in\Omega}}
\left|
\widehat R_{n,k}^{\mathrm{alt}}(\mathbf x)
-
\widehat R_{n,k}^{\mathrm{ref}}(\mathbf x)
\right|.
\tag{28}
\]

Here, \(N=1054\), \(\Omega\) denotes the set of valid grid points, and \(\mathrm{alt}\) and \(\mathrm{ref}\) denote the compensation residuals after subsequent-input replacement and under the original inputs, respectively. The values are reported on the 0--1 SIC scale.

\begin{table}[!t]
\centering
\footnotesize
\caption{Mean Absolute Changes in the Original Compensation Residuals Over Days 1--7 After Replacement of the Inputs for Days 8--14.}
\label{tab:4-5}
\resizebox{\columnwidth}{!}{%
\begin{tabular}{@{}lrrr@{}}
\toprule
\shortstack[l]{Subsequent-input\\replacement} & \shortstack[l]{H24 without temporal\\constraints} & \shortstack[l]{H24 with temporal\\constraints} & \shortstack[l]{H34 with temporal\\constraints} \\
\midrule
Zero forcing & 0.07404 & 0 & 0 \\
Noisy forcing & 0.09715 & 0 & 0 \\
All-zero inputs & 0.09365 & 0 & 0 \\
Repeated day-7 inputs & 0.01543 & 0 & 0 \\
Reversed tail & 0.01333 & 0 & 0 \\
\bottomrule
\end{tabular}%
}
\end{table}

\begin{figure}[!t]
\centering
\includegraphics[width=\columnwidth]{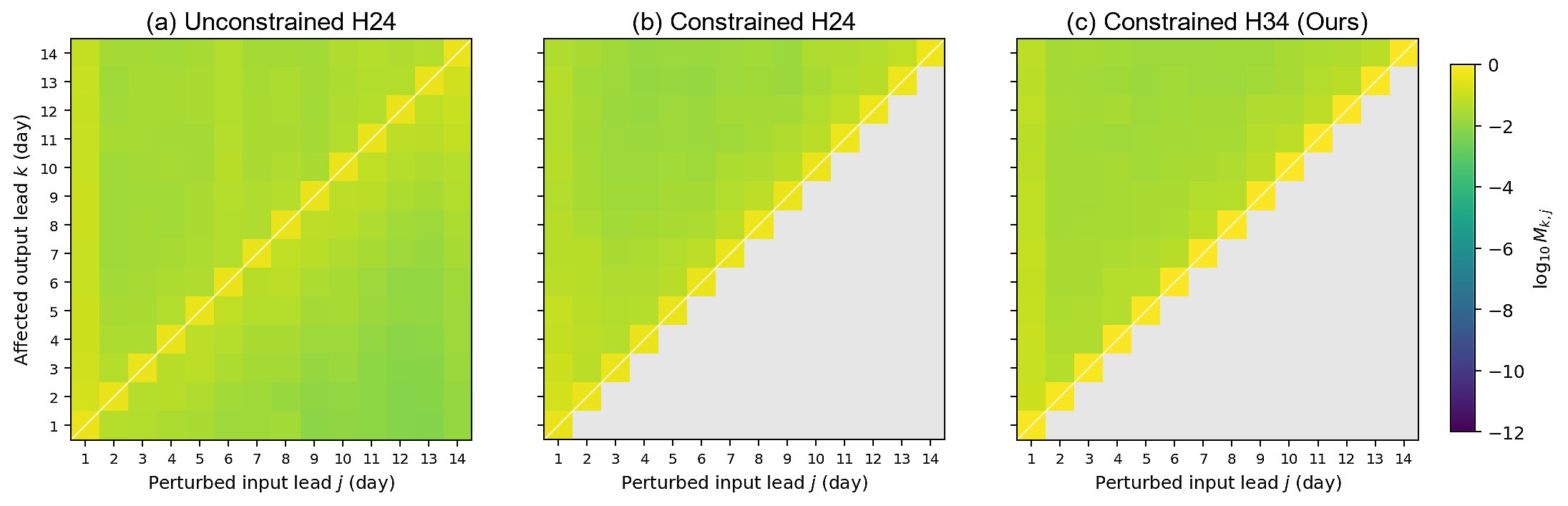}
\caption{Temporal influence matrices for H24 without temporal constraints, H24 with temporal constraints, and H34 with temporal constraints. The horizontal and vertical axes represent input lead \(j\) and output lead \(k\), respectively, with the vertical axis increasing upward. Gray denotes exact zeros, whereas the remaining colors represent \(\log_{10}M_{k,j}\).}
\label{fig:4-8}
\end{figure}

As shown in Table~\ref{tab:4-5}, both temporally constrained models maintain strict output-prefix invariance under all replacement schemes, whereas the first seven outputs of H24 without temporal constraints change in every case. In a further experiment, 12 initialization dates are sampled uniformly from each year, yielding 36 samples in total, and a fixed random perturbation is separately applied to each input lead in the standardized input space to construct a \(14\times14\) temporal influence matrix. The matrix element \(M_{k,j}\) denotes the mean absolute change in the compensation residual at output lead \(k\) after perturbation of input lead \(j\), averaged over all samples and valid grid points. As shown in Fig.~\ref{fig:4-8}, the earlier outputs of H24 without temporal constraints generally respond to later inputs, whereas the responses of H24 with temporal constraints and H34 with temporal constraints are zero at all positions satisfying \(j>k\). This result indicates that the prefix invariance of the temporally constrained models applies to arbitrary output lengths rather than only to the seven-day truncation setting.

\textbf{Accuracy Cost of the Temporal Constraint and Joint-Output Property.} Table~\ref{tab:4-6} summarizes the accuracy cost of the temporal constraint and the corresponding structural verification results. After the temporal constraint is imposed on the equal-capacity H24 model, the 14-day MAE increases from 0.018878 to 0.023708, indicating that exclusion of future inputs increases the learning difficulty. Increasing the hidden width to H34 reduces the MAE to 0.020235 while preserving strict output-prefix invariance.

\begin{table}[!t]
\centering
\scriptsize
\caption{Accuracy Cost of the Temporal Constraint and Structural Verification Results.}
\label{tab:4-6}
\resizebox{\columnwidth}{!}{%
\begin{tabular}{@{}
  l
  r
  r
  r
  r@{}}
\toprule
Structure & Parameter count & 14-day MAE & \shortstack[l]{Max 7-Day Residual-\\Prefix Drift} & \shortstack[l]{Max Future-Input\\Influence} \\
\midrule
No temporal constraint, H24\(^\dagger\) & 0.281 M & 0.018878 & 1.897592 & 0.132352 \\
Temporal constraint, H24 & 0.281 M & 0.023708 & 0 & 0 \\
Temporal constraint, H34 (Ours) & 0.556 M & \textbf{0.020235} & \textbf{0} & \textbf{0} \\
\bottomrule
\end{tabular}
}
\end{table}

\(^\dagger\) The lower error of the model without temporal constraints results from access to the complete 14-day input sequence. It is included only to quantify the accuracy cost of the temporal constraint and is excluded from the formal performance ranking of the temporally constrained models.

\subsubsection{Pathway-Specific Roles and Multiscale Contributions of Input Variables}

\textbf{Pathway Sensitivity of Input Variables}

To identify the pathways through which environmental variables contribute to the base evolution core, the target variables are separately replaced in the physical pathway, the compensation pathway, and both pathways. In the main text, each variable is replaced with its daily spatial climatology over the training period, preserving the seasonal and spatial background while removing daily variations. For physical-pathway replacement, the physical trajectory is recomputed; for compensation-pathway replacement, the physical trajectory is retained and only the direct forcing inputs to the compensation network are replaced. The complete experimental configuration is provided in Supplementary Materials B.3.

For variable \(v\), pathway \(p\), and evaluation metric \(E\), the relative error increase is defined as

\[
R_{v,p}^{(E)}
=
\frac{E_{v,p}-E_{\mathrm{full}}}
{E_{\mathrm{full}}}
\times 100\%,
\qquad
E\in\{\mathrm{MAE},\mathrm{IIEE}\},
\tag{29}
\]

where \(E_{v,p}\) and \(E_{\mathrm{full}}\) denote the errors after variable replacement and for the complete model, respectively.

The resulting pathway sensitivities are reported in Table~\ref{tab:4-7}.

\begin{table}[!t]
\centering
\scriptsize
\caption{Relative Error Increases for Different Variable Pathways Under Daily Spatial-Climatology Replacement (MAE / IIEE).}
\label{tab:4-7}
\resizebox{\columnwidth}{!}{%
\begin{tabular}{@{}lrrr@{}}
\toprule
Variable & Physical pathway & Compensation pathway & Both pathways \\
\midrule
Sea ice velocity & +17.24\% / +20.41\% & +29.63\% / +57.98\% & +53.01\% / +79.80\% \\
Sea surface temperature & +27.05\% / +15.66\% & +3.12\% / +5.10\% & +25.66\% / +12.72\% \\
2-m air temperature & +6.41\% / +14.72\% & +0.42\% / +0.98\% & +6.32\% / +13.13\% \\
Net heat flux & +13.15\% / +4.03\% & +1.59\% / +0.54\% & +14.88\% / +3.26\% \\
\bottomrule
\end{tabular}
}
\end{table}

Note: Sea ice velocity comprises the two components \(u_i\) and \(v_i\).

As shown in Table~\ref{tab:4-7}, sea ice velocity contributes through both the physical and compensation pathways, with greater sensitivity in the compensation pathway. Sea surface temperature, 2-m air temperature, and net heat flux contribute to state evolution primarily through the physical pathway, indicating a pathway-specific division in which sea ice velocity acts through both pathways whereas thermodynamic variables are primarily represented by the physical pathway.

The IIEE increase induced by sea ice velocity replacement is substantially larger than the MAE increase, indicating its dominant influence on ice-edge position and spatial transport. The stronger effect of net heat flux on MAE reflects its primary role in regulating local freezing and melting. The response to simultaneous replacement in both pathways is not equal to the sum of the responses from the individual pathways, indicating nonlinear interactions between the physical baseline and compensation network. Therefore, these results characterize pathway sensitivity rather than additive geophysical causal contributions.

\textbf{Multiscale Response.} To examine the retention and accumulation of variable dependence during continuous integration, the target variable is simultaneously replaced with its daily spatial climatology over the training period in both the physical and compensation pathways, followed by a daily Lead-1 closed-loop rollout. For variable \(v\), the cumulative relative error increase over integration horizon \(H\) is defined as

\[
R_v^{(H,E)}
=
\frac{
E_v^{(1:H)}-E_{\mathrm{full}}^{(1:H)}
}{
E_{\mathrm{full}}^{(1:H)}
}
\times 100\%,
\qquad
E\in\{\mathrm{MAE},\mathrm{IIEE}\},
\tag{30}
\]

where \(E_v^{(1:H)}\) and \(E_{\mathrm{full}}^{(1:H)}\) denote the cumulative errors after variable replacement and for the complete model, respectively. The complete experimental configuration is provided in Supplementary Materials B.3.

\begin{figure}[!t]
\centering
\includegraphics[width=\columnwidth]{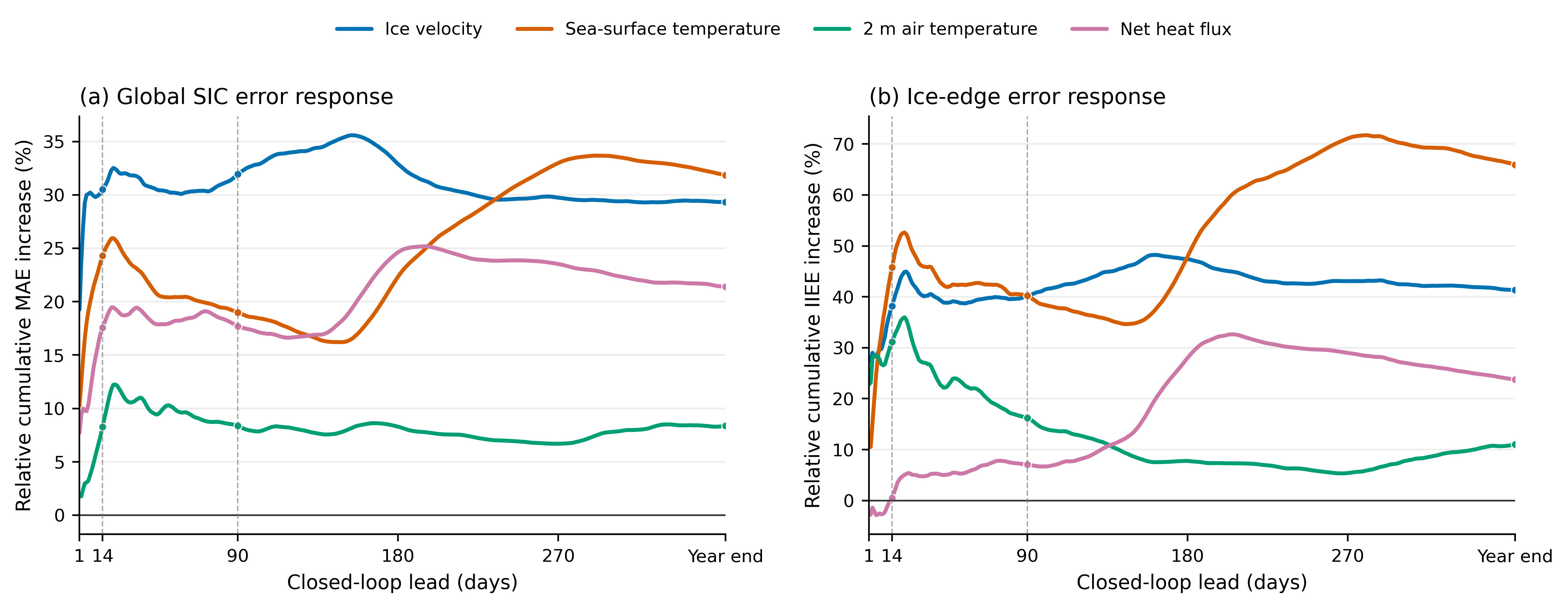}
\caption{Cumulative relative increases in MAE and IIEE for different variables under daily spatial-climatology replacement. Dashed lines mark days 14 and 90, and points denote the results at day 14, day 90, and the end of the year.}
\label{fig:4-9}
\end{figure}

The resulting multiscale responses are shown in Fig.~\ref{fig:4-9}.

As shown in Fig.~\ref{fig:4-9}, the MAE and IIEE increases associated with sea ice velocity remain approximately 30\% and 38--41\%, respectively, across different timescales, indicating its persistent control over spatial transport and ice-edge position. The effect of sea surface temperature increases markedly during the melt season, with end-of-year MAE and IIEE increases reaching 31.83\% and 65.93\%, respectively, reflecting its long-term influence on the thermodynamic state.

The IIEE increase associated with net heat flux rises from 0.52\% at day 14 to 23.75\% at the end of the year, demonstrating a pronounced cumulative effect. The MAE increase associated with 2-m air temperature remains approximately 8\%, whereas its IIEE increase declines from 31.16\% to 11.02\%, indicating a strong state dependence. Overall, the model exhibits a multiscale variable-dependence structure characterized by stable transport from sea ice velocity, seasonal regulation by sea surface temperature, progressive accumulation from net heat flux, and conditional supplementation from 2-m air temperature. These results characterize model sensitivity rather than additive geophysical causal contributions.

\textbf{Annual Sea Ice Area Response.} To determine whether the multiscale error responses correspond to changes in annual total ice amount, the SIA evolution after variable replacement is further examined.

\begin{figure}[!t]
\centering
\includegraphics[width=\columnwidth]{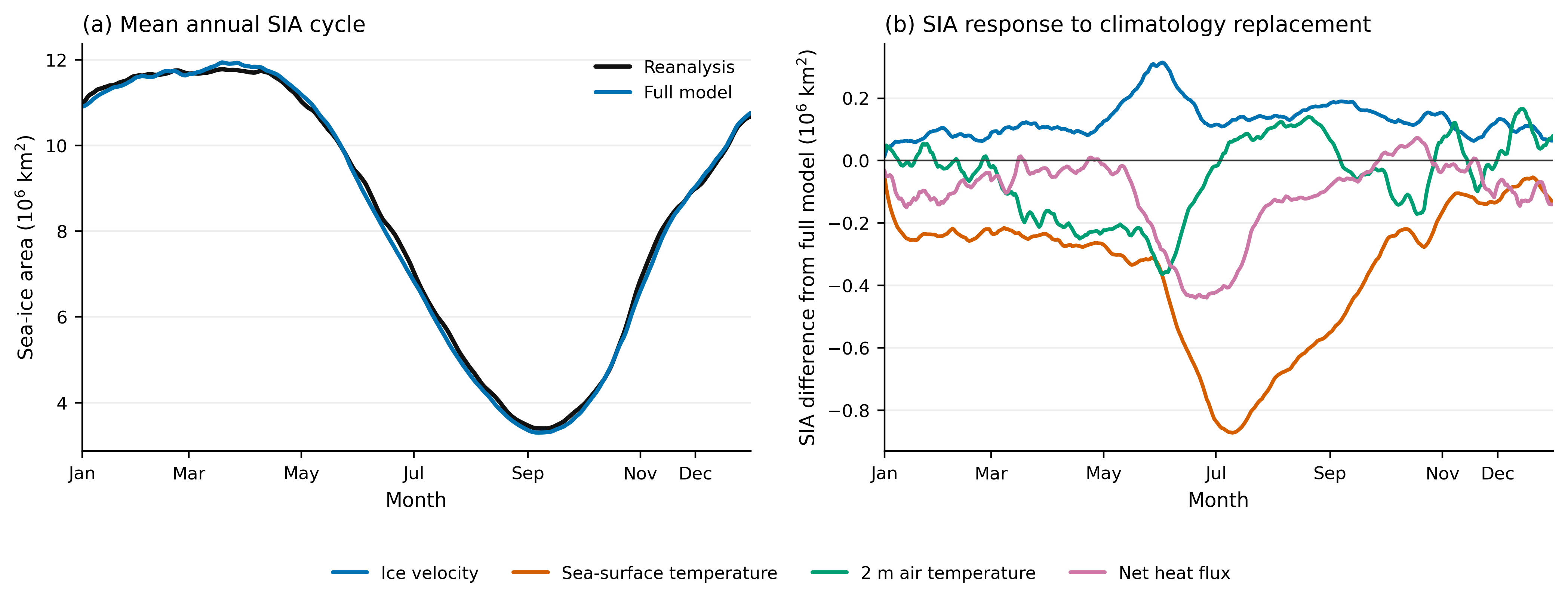}
\caption{Annual SIA response under daily spatial-climatology replacement: (a) three-year mean seasonal cycles of the complete model and the reanalysis reference; (b) SIA differences between the variable-replacement results and the complete model.}
\label{fig:4-10}
\end{figure}

The annual SIA responses are shown in Fig.~\ref{fig:4-10}.

As shown in Fig.~\ref{fig:4-10}(a), the complete model stably preserves the annual freezing--melting cycle. As shown in Fig.~\ref{fig:4-10}(b), the influence of sea surface temperature is concentrated mainly in the melt season, net heat flux exhibits a seasonally accumulated effect, sea ice velocity induces a persistent SIA shift, and the response to 2-m air temperature varies seasonally. Together with Fig.~\ref{fig:4-9}, these results indicate a variable-dependence structure characterized by stable transport from sea ice velocity, thermodynamic-state representation by sea surface temperature, seasonal source--sink regulation by net heat flux, and conditional supplementation from 2-m air temperature. These results characterize model sensitivity rather than additive geophysical causal contributions.

\subsection{Forecast-Forced Prediction and Variable-Semantic Adaptation}

After validation of the base evolution core, all core parameters are frozen, and only the newly introduced forecast-domain adaptation modules are trained to map observational initial conditions and forecast forcing into input semantics compatible with the evolution core. The forecasting task is formulated as

\[
A_T^{\mathrm{OSI}},\ 
\mathbf F_{T:T+9}^{\mathrm{GFS/TOPAZ5}}
\longrightarrow
\widehat A_{T+1:T+10}^{\mathrm{OSI}}.
\tag{31}
\]

Under the lead-prefix constraint, the day-\(k\) prediction depends only on forcing through day \(k\), and subsequent forcing does not affect existing outputs; forecasts can therefore follow the available forcing horizon.

Here, the initial SIC and verification targets are obtained from OSI SAF, the 2-m air temperature and net heat flux from GFS, and the sea ice velocity and sea surface temperature from TOPAZ5. The adaptation modules are trained on 366 initialization dates from 1 September 2023 to 31 August 2024 and evaluated over ten days on 336 valid initialization dates from 18 September 2024 to 1 September 2025.

\subsubsection{Forecast-Domain Transfer and Overall Prediction Performance}

This section examines whether the frozen reanalysis-domain evolution core can be transferred to the forecast domain through lightweight adaptation. Accordingly, the overall performance of direct transfer and progressive adaptation is first compared, followed by lead-wise analyses of MAE and IIEE and an equal-parameter-budget experiment that determines whether the performance gains arise from the adaptation structure.

Without adaptation, directly replacing reanalysis inputs with observational SIC and forecast forcing increases MAE to 0.125973 and produces a pronounced negative bias, indicating that corresponding variables in the reanalysis and forecast domains exhibit distinct data distributions and error semantics. Table~\ref{tab:4-8} reports the results obtained by progressively introducing the forecast-domain adaptation mechanisms.

\begin{table*}[!t]
\centering
\scriptsize
\caption{Ten-Day Mean Performance Under Progressive Forecast-Domain Adaptation. IIEE and SIAE Are Reported in \(10^6\ \mathrm{km}^2\).}
\label{tab:4-8}
\begin{tabular}{@{}
  >{\raggedright\arraybackslash}p{(\textwidth - 10\tabcolsep) * \real{0.30}}
  >{\raggedleft\arraybackslash}p{(\textwidth - 10\tabcolsep) * \real{0.14}}
  >{\raggedleft\arraybackslash}p{(\textwidth - 10\tabcolsep) * \real{0.14}}
  >{\raggedleft\arraybackslash}p{(\textwidth - 10\tabcolsep) * \real{0.14}}
  >{\raggedleft\arraybackslash}p{(\textwidth - 10\tabcolsep) * \real{0.14}}
  >{\raggedleft\arraybackslash}p{(\textwidth - 10\tabcolsep) * \real{0.14}}@{}}
\toprule
Method & MAE & RMSE & Bias & IIEE & SIAE \\
\midrule
Persistence & 0.033642 & 0.090711 & \textbf{+0.000137} & 0.508404 & 0.310133 \\
Direct replacement & 0.125973 & 0.208540 & -0.093977 & 1.245631 & 1.973573 \\
Initial-domain adaptation & 0.038366 & 0.089657 & -0.010703 & 0.476152 & 0.240168 \\
Variable-semantic adaptation & 0.032689 & 0.078341 & -0.001965 & 0.436107 & 0.125756 \\
Dynamical--thermodynamic refinement (Ours) & \textbf{0.032401} & \textbf{0.075930} & +0.000211 & \textbf{0.428574} & \textbf{0.119582} \\
\bottomrule
\end{tabular}
\end{table*}

Initial-domain adaptation substantially reduces the systematic bias induced by direct replacement, while variable-semantic adaptation further improves grid-point accuracy, ice-edge position, and sea ice area. Dynamical--thermodynamic refinement further achieves the lowest MAE, RMSE, IIEE, and SIAE. Relative to Persistence, the final model reduces MAE, RMSE, IIEE, and SIAE by 3.69\%, 16.30\%, 15.70\%, and 61.44\%, respectively.

\begin{figure}[!t]
\centering
\includegraphics[width=\columnwidth]{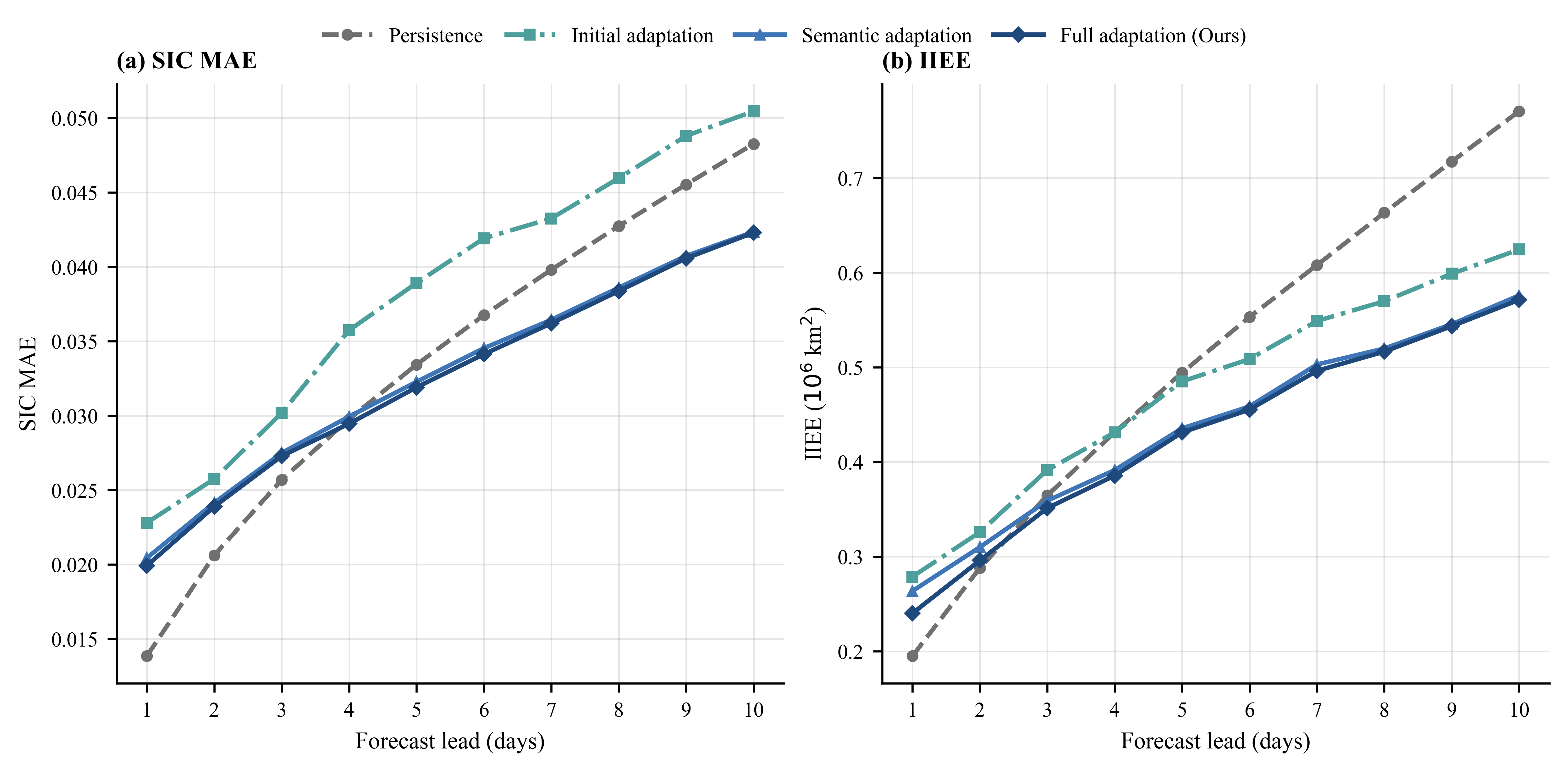}
\caption{Lead-wise performance of different adaptation models: (a) SIC MAE over days 1--10; (b) IIEE over days 1--10.}
\label{fig:4-11}
\end{figure}

Lead-wise performance is shown in Fig.~\ref{fig:4-11}.

As shown in Fig.~\ref{fig:4-11}, Persistence maintains a lower MAE during the first three days, whereas the final model becomes superior from day 4 onward. For IIEE, the final model outperforms Persistence from day 3 onward. On day 10, the final model reduces MAE and IIEE by 12.36\% and 25.86\%, respectively, relative to Persistence. This result indicates that informative signals from forecast forcing become increasingly effective with lead time and yield more pronounced improvements in ice-edge position.

To distinguish the effect of adaptation structure from that of parameter count, a further comparison is conducted under identical training samples, training epochs, and approximately equal numbers of additional parameters. Each comparison network adopts a lead-shared two-layer CNN that concatenates the observational initial condition with forecast forcing at each lead for unified correction, whereas the proposed approach decomposes adaptation according to the distinct semantics of the state, vector sea ice velocity, scalar thermodynamic forcing, and evolution response. All models retain their respective reanalysis-domain backbones as fixed components and optimize only the newly introduced forecast-domain adaptation parameters.

\begin{figure}[!t]
\centering
\includegraphics[width=\columnwidth]{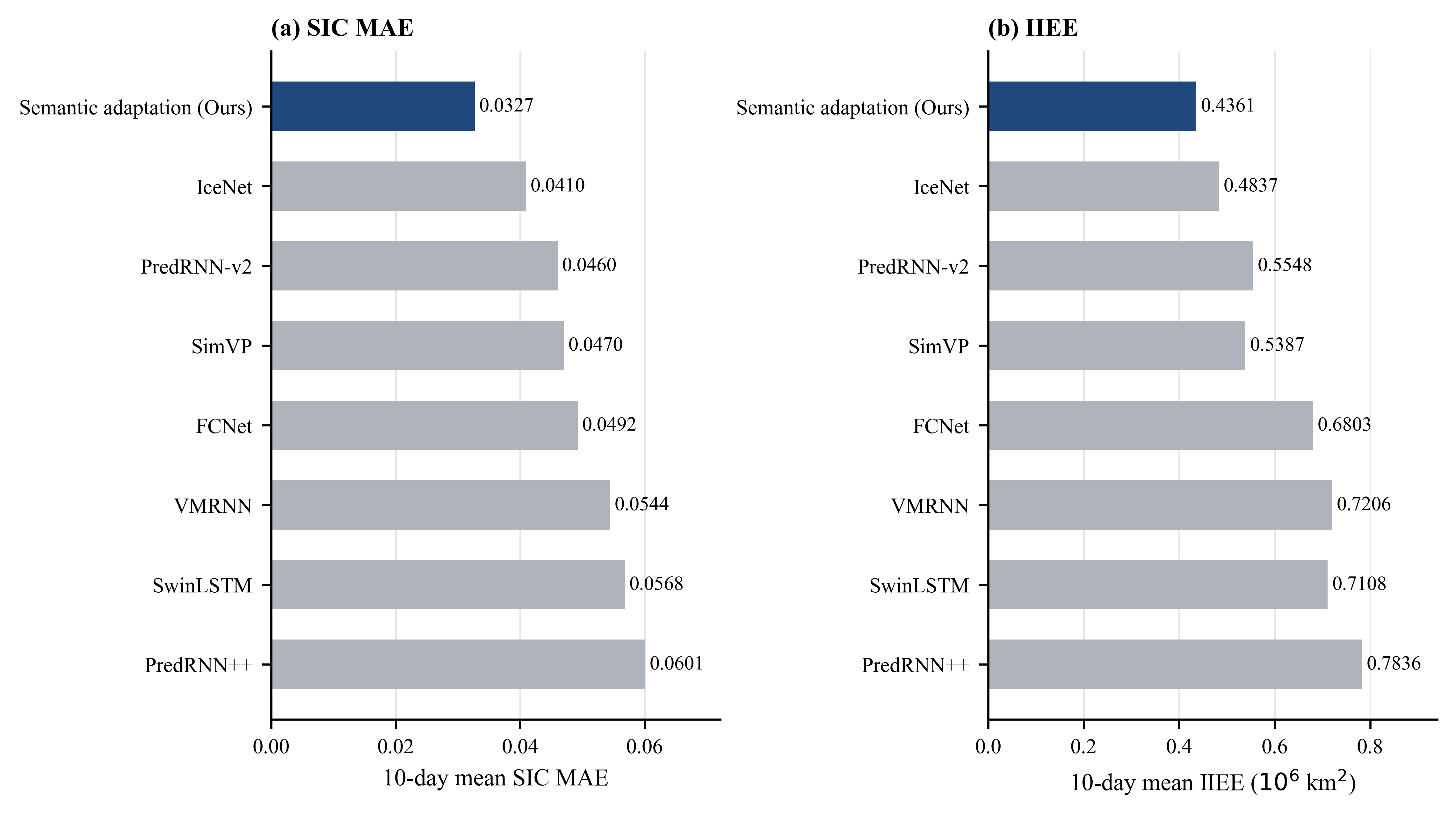}
\caption{Forecast-domain adaptation under an equal parameter budget: (a) ten-day mean SIC MAE; (b) ten-day mean IIEE. Ours denotes the variable-semantic adaptation model under the equal-budget setting.}
\label{fig:4-12}
\end{figure}

The resulting equal-parameter-budget comparison is shown in Fig.~\ref{fig:4-12}.

As shown in Fig.~\ref{fig:4-12}, under approximately equal numbers of additional parameters, variable-semantic adaptation attains an MAE of 0.032689 and an IIEE of \(0.436107\times10^6\ \mathrm{km}^2\), both lower than those of all comparison models, demonstrating consistent advantages in overall accuracy and ice-edge simulation. This result indicates that forecast-domain performance recovery arises primarily from an adaptation structure aligned with variable-specific error semantics rather than from additional parameter capacity.

Overall, although the frozen reanalysis-domain evolution core cannot be transferred to the forecast domain through direct input replacement, lightweight variable-semantic adaptation restores its overall prediction accuracy and ice-edge simulation capability.

\subsubsection{Analysis of Adaptation Mechanisms and Variable Contributions}

\textbf{Functional Division of the Adaptation Modules.} To identify the sources of performance gains from variable-semantic adaptation, sea-ice-velocity correction, flux correction, pathway gating, and the semantic-response structure are activated separately while the model parameters remain fixed. The performance of the different module configurations is reported in Table~\ref{tab:4-9}.

\begin{table*}[!t]
\centering
\scriptsize
\caption{Performance of Different Adaptation-Module Configurations. IIEE and SIAE Are Reported in \(10^6\ \mathrm{km}^2\).}
\label{tab:4-9}
\begin{tabular}{@{}
  >{\raggedright\arraybackslash}p{(\textwidth - 10\tabcolsep) * \real{0.25}}
  >{\raggedleft\arraybackslash}p{(\textwidth - 10\tabcolsep) * \real{0.15}}
  >{\raggedleft\arraybackslash}p{(\textwidth - 10\tabcolsep) * \real{0.15}}
  >{\raggedleft\arraybackslash}p{(\textwidth - 10\tabcolsep) * \real{0.15}}
  >{\raggedleft\arraybackslash}p{(\textwidth - 10\tabcolsep) * \real{0.15}}
  >{\raggedleft\arraybackslash}p{(\textwidth - 10\tabcolsep) * \real{0.15}}@{}}
\toprule
Adaptation configuration & MAE & RMSE & Bias & IIEE & SIAE \\
\midrule
Initial adaptation & 0.038366 & 0.089657 & -0.010703 & 0.476152 & 0.240168 \\
Sea-ice-velocity correction only & 0.037833 & 0.087556 & -0.011037 & 0.465355 & 0.247052 \\
Flux correction only & 0.038000 & 0.089564 & -0.008301 & 0.471681 & 0.224860 \\
Sea-ice velocity + flux + gating & 0.037757 & 0.087717 & -0.008534 & 0.455445 & 0.244399 \\
Complete semantic structure & \textbf{0.032689} & \textbf{0.078341} & \textbf{-0.001965} & \textbf{0.436107} & \textbf{0.125756} \\
\bottomrule
\end{tabular}
\end{table*}

Sea-ice-velocity correction primarily improves overall error and ice-edge position, whereas flux correction is more effective for systematic bias and sea ice area. The combined pathway-gating configuration further reduces MAE, RMSE, and IIEE, although the overall improvement remains limited. The complete semantic structure achieves the best results across all metrics and substantially reduces the negative bias and area error.

\textbf{Functional Diagnostics of Forecast-Domain Process-Refinement Branches.} To analyze the effects of dynamical and thermodynamic process corrections on forecast performance, the reanalysis-domain evolution core and variable-semantic adaptation backbone are fixed, while dynamical, freezing, melting, and dynamical--thermodynamic interaction branches are jointly trained on the base predictions. Each branch receives the physical trajectories and diagnostics generated by the fixed backbone and corrects the base prediction additively:

\[
\begin{aligned}
\widehat A
&=\Pi_{[0,1]}\!\biggl(
\widehat A_{\mathrm{base}}+c_{\mathrm d}\Delta A_{\mathrm{dyn}}\\
&\quad+c_{\mathrm f}\Delta A_{\mathrm{frz}}
 +c_{\mathrm m}\Delta A_{\mathrm{melt}}\\
&\quad+c_{\mathrm i}\Delta A_{\mathrm{int}}
\biggr),
\end{aligned}
\tag{32}
\]

Here, \(\widehat A_{\mathrm{base}}\) denotes the base prediction from the fixed backbone; \(\Delta A_{\mathrm{dyn}}\), \(\Delta A_{\mathrm{frz}}\), \(\Delta A_{\mathrm{melt}}\), and \(\Delta A_{\mathrm{int}}\) denote the dynamical, freezing, melting, and interaction corrections, respectively; and \(c_{\mathrm d},c_{\mathrm f},c_{\mathrm m},c_{\mathrm i}\in\{0,1\}\) determine whether the corresponding branches contribute to the output. Table~\ref{tab:4-10} switches these branches under the same jointly trained parameter set to diagnose the actual contribution of each correction term within the complete model.

\begin{table*}[!t]
\centering
\scriptsize
\caption{Performance of Different Process-Refinement Branch Configurations Under Fixed Jointly Trained Parameters. IIEE and SIAE Are Reported in \(10^6\ \mathrm{km}^2\).}
\label{tab:4-10}
\begin{tabular}{@{}
  >{\raggedright\arraybackslash}p{(\textwidth - 10\tabcolsep) * \real{0.30}}
  >{\raggedleft\arraybackslash}p{(\textwidth - 10\tabcolsep) * \real{0.14}}
  >{\raggedleft\arraybackslash}p{(\textwidth - 10\tabcolsep) * \real{0.14}}
  >{\raggedleft\arraybackslash}p{(\textwidth - 10\tabcolsep) * \real{0.14}}
  >{\raggedleft\arraybackslash}p{(\textwidth - 10\tabcolsep) * \real{0.14}}
  >{\raggedleft\arraybackslash}p{(\textwidth - 10\tabcolsep) * \real{0.14}}@{}}
\toprule
Refinement structure & MAE & RMSE & Bias & IIEE & SIAE \\
\midrule
Variable-semantic adaptation baseline & 0.032689 & 0.078341 & -0.001965 & 0.436107 & 0.125756 \\
Dynamical branch only & 0.032823 & 0.077184 & -0.003126 & 0.432103 & 0.125337 \\
Freezing/melting branches only & 0.033264 & 0.077404 & +0.001256 & 0.432064 & 0.127710 \\
Dynamical + freezing/melting & \textbf{0.032401} & 0.075930 & \textbf{+0.000211} & \textbf{0.428574} & 0.119582 \\
Dynamical + freezing/melting + interaction & 0.032570 & \textbf{0.075874} & -0.000587 & 0.429033 & \textbf{0.115946} \\
\bottomrule
\end{tabular}
\end{table*}

Activating either the dynamical branch or the freezing/melting branches individually reduces RMSE and IIEE, but the former increases the negative bias, whereas the latter introduces a positive bias and increases SIAE. Joint activation of the dynamical and freezing/melting branches produces complementary error corrections and improves all major metrics relative to the variable-semantic adaptation baseline. Although the explicit interaction branch further reduces RMSE and SIAE, it slightly degrades MAE, Bias, and IIEE; therefore, the constrained additive configuration without an explicit interaction term is adopted. These comparisons characterize the functional contributions of the branches within the jointly trained model rather than the performance upper bounds of independently retrained reduced structures.

\textbf{Effective Exploitation of Future Forecast Information}

To assess whether the model genuinely incorporates future forecast forcing, each variable is separately subjected to same-month temporal misalignment and sequence persistence. Same-month temporal misalignment preserves the seasonal distribution while removing event correspondence, whereas sequence persistence fixes every forecast lead to the field at the first forecast lead, thereby eliminating future variation. All other inputs and model parameters remain unchanged.

\begin{figure}[!t]
\centering
\includegraphics[width=\columnwidth]{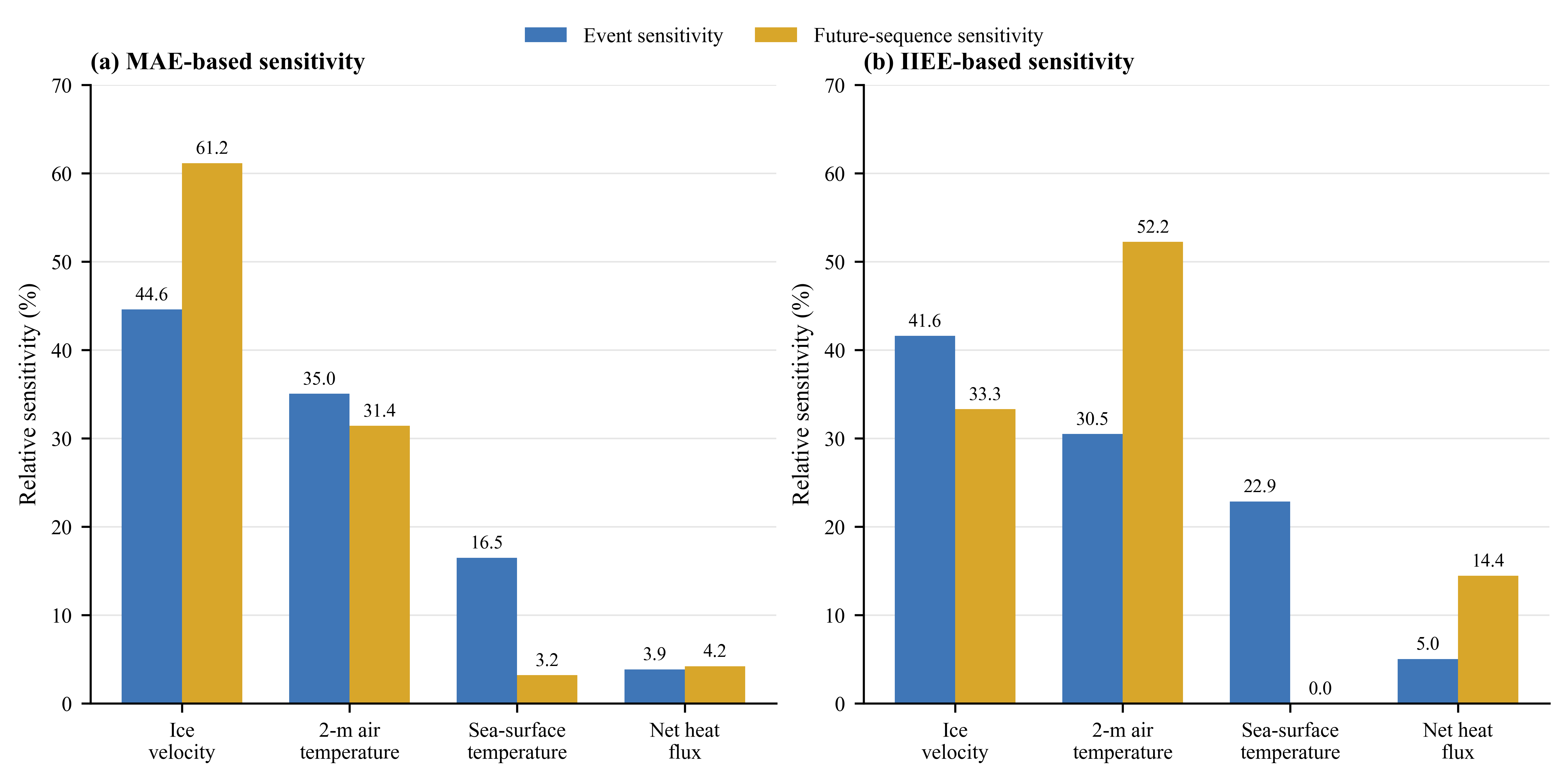}
\caption{Relative sensitivity to future forecast information: (a) based on MAE; (b) based on IIEE. Percentages are normalized only over the positive error increases of the four external forcing variables and exclude the initial SIC.}
\label{fig:4-13}
\end{figure}

The resulting sensitivity is shown in Fig.~\ref{fig:4-13}.

As shown in Fig.~\ref{fig:4-13}, the overall SIC field depends primarily on event correspondence and future sequence information from sea ice velocity and 2-m air temperature. The ice edge is most sensitive to event information in sea ice velocity, whereas its response to future variation depends primarily on 2-m air temperature and sea ice velocity. Sea surface temperature primarily provides a slowly varying thermodynamic-state background, and its persistence does not increase IIEE. Net heat flux has a weaker effect on the overall field but contributes to future ice-edge changes.

These results indicate that the model relies not only on the observed initial SIC but also incorporates event correspondence and future variation contained in the forecast forcing. Therefore, forecast performance cannot be explained solely by sea-ice persistence; it also arises from the selective exploitation of future forecast information through variable-semantic adaptation. These results characterize model sensitivity rather than geophysical causal contributions.

\section{Conclusions and Outlook}

This study proposes a reanalysis-forecast dual-domain decoupled deep learning framework for constructing a daily SIC evolution model with an explicit evolution structure and verifiable process roles, and for applying it to practical forecasting. The framework learns a base evolution core from reanalysis data and represents daily evolution relationships through physics guidance and temporal constraints; in the forecasting stage, this core is kept fixed, while lightweight variable-semantic adaptation alone addresses inter-domain differences induced by observational initial conditions and forecast forcing.

The experimental results show that the base evolution core can accurately and stably simulate SIC evolution across different time scales. Mechanistic analyses verify the functional division between physical evolution and data-driven compensation, as well as the roles of the ice-mass-aware mechanism and temporal constraints in evolution stability. On this basis, variable-semantic adaptation effectively mitigates performance degradation when the model is transferred from the reanalysis domain to the forecast domain, enabling the base evolution core to support sea ice prediction under forecast-forcing conditions. These results indicate that sea ice evolution relationships and forecast-domain errors can be modeled by mutually independent parameter systems, thereby balancing the stability of the evolution structure and the adaptability required for practical forecasting.

Current research remains constrained by data coverage, forecast products, and deterministic modeling. Future work will conduct independent validation across different years and forecast products, investigate uncertainty propagation arising from observational and forecast errors, and further improve the computational efficiency of joint multi-lead prediction and the model's cross-domain generalization capability.

\end{document}